\documentclass{aa}  

\usepackage{graphicx}
\usepackage{txfonts}
\usepackage{hyperref}
\hypersetup{
    colorlinks=true,
    linkcolor=blue,
    filecolor=magenta,      
    urlcolor=blue,
    citecolor=blue,
    } 
\begin{document}

   \title{Foreground and internal free--free absorption in particle-accelerating colliding-wind binaries}

\titlerunning{FFA in PACWBs}
\authorrunning{Tasseroul et al.}
   \subtitle{Insights from the radio emission of WR\,147}

   \author{M. Tasseroul
          \inst{1}
          \and
          M. De Becker\inst{1}
          \and
          A. B. Blanco\inst{1}
          \and
          P. Benaglia\inst{2}
          \and
          S. del Palacio\inst{3}
          }

   \institute{Space Sciences, Technologies and Astrophysics Research (STAR) Institute, University of Liège, Quartier Agora, 19c,
Allée du 6 Août, B5c, 4000 Sart Tilman,
               Belgium \\
              \email{Mathilde.Tasseroul@uliege.be}
              \and
              Instituto Argentino de Radioastronomia (CONICET; CICPBA; UNLP), C.C. No 5, 1894, Villa Elisa, Argentina
              \and
              Department of Space, Earth and Environment, Chalmers University of Technology, SE-412 96 Gothenburg, Sweden}

   \date{Received September 15, 1996; accepted March 16, 1997}

  \abstract  
   {Radio emission from massive binary systems is generally of composite nature, showing both a thermal emission component from the winds and a non-thermal component from relativistic electrons accelerated in the colliding-wind region. Understanding the processes ruling their radio spectrum is essential to investigate the role of these objects in the production of non-thermal particle populations in our galaxy.}
   {Our objective is to explore how the processes at work in particle-accelerating colliding-wind binaries (PACWBs) alter their spectral energy distribution, following a simple phenomenological description. We focus mainly on the role of free--free absorption (FFA) at low frequencies. Our intention is to use WR\,147 as a test case, before tentatively extrapolating to a more generic behaviour.}
   {We processed recent Karl G. Jansky Very Large Array data, optimised for spectral analysis, combined with older measurements published in the literature at other frequencies. We analysed the radio spectrum considering both a more classical foreground free--free absorption (f-FFA) model and, for the first time, an internal free--free absorption (i-FFA) model.}
   {Our results show that the f-FFA model does not reproduce the spectral energy distribution of WR\,147 at low frequencies. The i-FFA model is more efficient in providing a more complete description of the SED down to 610 MHz. This model is the only one to account for a change in the spectral index at low frequencies without any exponential drop in flux, as predicted by the f-FFA model. In addition, the upper limit at 150\,MHz shows that two turnovers occur in the radio spectrum of WR\,147, suggesting the effect of both i-FFA and f-FFA is seen in two regions of the spectrum.}
   {The radio spectrum at low frequencies for very long period systems may display some internally attenuated synchrotron emission, without necessarily being suppressed by a steep exponential cut-off. We propose a generic spectral energy distribution for FFA-affected radio spectra of colliding-wind massive binaries, where the overall spectral shape can be expressed in terms of the relative importance of f-FFA and i-FFA. We also comment on the observational consequences of this generic behaviour in the broader context of the full class of PACWBs.}


   \maketitle

\section{Introduction}\label{intro}
The radio investigation of massive stars (O-type and Wolf-Rayet (WR)), revealed their ability to act both as thermal and non-thermal emitters \citep{ABC,ABCT,BAC,WhBe}. Although thermal emission is attributed to bremsstrahlung arising from the hot, ionised winds \citep{WB,PF}, the non-thermal component is known to be due to synchrotron radiation \citep{Wh}. The latter radio contribution requires the prior acceleration of relativistic electrons, most likely occurring through Diffusive Shock Acceleration (DSA, \citealt{Drury1983}). In the context of massive stars, this process takes place in the vicinity of the high Mach number shocks produced by the colliding hypersonic winds in binary, or higher multiplicity systems. Particle acceleration and non-thermal emission processes from such systems have been the topic of numerous studies in the past few decades \citep[e.g.][]{EU,Doug,Pit2,RPR2006,debeckerreview,Benaglia2025}. The subset of colliding-wind binaries known to accelerate particles is referred to as particle-accelerating colliding-wind binaries (PACWBs). The catalogue of PACWBs includes more than 50 systems \citep{catapacwb}\footnote{The updated version of the catalogue is available at \url{https://www.astro.uliege.be/~debecker/pacwb/}.}.

One key question worth investigating about these objects is the fraction of PACWBs among colliding-wind systems \citep[e.g.][]{DeBecker2017}. Although massive stars are mostly found in multiple systems \citep{Offner2023}, evidence of particle acceleration is not always easy to provide. As shown by \citet{catapacwb}, the most efficient tracer for particle acceleration so far is synchrotron radio emission. However, significant biases inhibit the detectability of synchrotron emission. The main reason is the occurrence of free--free absorption (FFA). Thermal plasmas are quite efficient at absorbing radio photons, in particular dense stellar winds. This is well illustrated, for example, by several seminal works on this topic, e.g. \citet{Doug,Doug140,Pit2}. If the synchrotron emission region is sufficiently embedded in dense wind material or is located right behind a dense wind, the synchrotron emission component may be (almost) fully absorbed. FFA is pointed out as the most likely cause for a lack of detection of non-thermal radio emission from several massive star systems, at least in some parts of their orbit \citep{DeBecker2019,Arora2021,Blanco2024}. 

The lack of information on the radio emission from many systems prevents us from achieving an insightful view of their behaviour. It is therefore important to study in some detail specific targets that have benefited from extensive observing campaigns and that are bright enough to provide us with higher quality measurements over several radio bands. Among the few targets that comply with these criteria, we decided to focus on WR\,147. This system has been a test case for modelling studies \citep{Doug}, in addition to being a target for numerous observation campaigns \citep{Contreras1999,williamswr147,skinner-wr147,wr147setia,Benaglia2020}. Our intention is to make use of unpublished data obtained with the Karl G. Jansky Very Large Array (VLA) to achieve a more accurate view of the radio spectral energy distribution (SED) of WR\,147 through a detailed intra-band analysis. We thus report on a SED analysis based on a broader and denser spectral coverage as compared to previous studies. A second important motivation for this work is to learn from the case of WR\,147 in order to improve our phenomenological description of the radio SED of PACWBs in general, especially with regard to the strong influence of FFA on the spectral morphology at lower frequencies. In other words, we aim at exploring how a simple and relevant description of their radio emission can be achieved in the absence of full hydro-radiative modelling of these complex systems.

The paper is organised as follows. A synthesis of what is known about the radio behaviour of the target is described in Sect.\,\ref{target}. The data and their processing are presented in Sect.\,\ref{obs}. In Sect.\,\ref{spec}, we proceed with a first detailed discussion focusing on WR\,147 before addressing some important implications for the investigation of the PACWB class of objects as a whole in Sect.\,\ref{disc}. We finally conclude in Sect.\,\ref{conc}.

\section{The target: WR\,147}\label{target}
WR\,147 (RA, DEC = 20$^\mathrm{h}$\,36$^\mathrm{m}$\,43.57$^\mathrm{s}$,
$+40^{\circ}\,21\arcmin\,7.5\arcsec$) is made of a WN8 primary with an early B-type or O8-9\,V-III secondary \citep{Niemela1998}. Although previously claimed to be the second closest WR with a distance of approximately 630\,pc \citep{Churchwell1992}, the Gaia DR3 distance is reported to be about 2\,kpc \citep{GaiaDR3}, with a revision to 1.7\,kpc by \citet{Crowther2023}. 

This system is one of the few to have been spatially resolved at radio frequencies \citep{Churchwell1992,williamswr147}, providing compelling evidence for strong synchrotron emission arising from the colliding-wind region (the northern component, referred to as WR\,147N) spatially shifted from the thermal emission from the WR wind (WR\,147S). The angular offset is about 0.6$''$. WR\,147 is among the top three brightest synchrotron-emitting PACWBs known to date, along with WR\,146 \citep{Dougwr146} and Apep \citep{Marcote2021}. The orbital period is not determined, but such a separation implies a value of several thousand years.

The analysis of its radio SED has been addressed in particular by \citet{skinner-wr147}, \citet{wr147setia} and \citet{Benaglia2020}. These studies focused on low angular resolution interferometric data. As the thermal and non-thermal emission components were not resolved in these measurements, these studies considered the fit of a composite emission spectrum. In order to favour the convergence of the fit, the thermal component was set to comply with the canonical optically thick free--free emission \citep{WB}, with a fixed spectral index $\alpha_\mathrm{T} = 0.6$ (for a flux density defined as $S_\nu^\mathrm{T} \propto \nu^{\alpha_\mathrm{T}}$). Within the uncertainties on the radio measurements, reasonably good results were obtained for what we will call a composite spectrum with foreground free--free absorption (hereafter f-FFA) of the non-thermal emission. In the framework of this model, the measured flux density ($S_\nu^\mathrm{obs}$) at a given frequency is represented as follows,
\begin{equation}\label{eq:SEDfFFA}
S_\nu^\mathrm{obs} = A\,\nu^{\alpha_\mathrm{T}} + B\,\nu^{-\alpha_\mathrm{NT}}\,e^{-\tau_\nu},
\end{equation}

\noindent where $\alpha_\mathrm{T}$ and $\alpha_\mathrm{NT}$ are the thermal and non-thermal spectral indices, and $A$ and $B$ are normalisation parameters of both components. $\tau_\nu$ is the optical depth that scales as a function of frequency as $\tau_\nu = (\nu/\nu_\mathrm{f-FFA})^{-2.1}$, where $\nu_\mathrm{f-FFA}$ is the turnover frequency for foreground FFA (where $\tau_\nu = 1$). The power appearing in the expression of the optical depth includes the dependence on the frequency of the Gaunt factor in the expression of the free--free absorption coefficient. We clarify that Eq.\,\ref{eq:SEDfFFA} explicitly considers that the thermal emission from both stellar winds is represented by the first term. In the specific case of a WR + OB binary, the thermal emission is completely dominated by the WR wind.

However, we identify a few points of improvement with respect to the aforementioned studies that are worth investigating:
\begin{enumerate}
    \item[-] Setting $\alpha_\mathrm{T}$ to 0.6 is certainly fairly adequate for O-type stars, but for WR winds the situation may be different. Radio measurements have shown that the thermal spectrum of WR winds is generally steeper than predicted by the canonical law. \citet{Nugis1998} reported on thermal spectral indices as steep as 0.7 or even 0.8, very likely as a result of the strong clumpiness of WR winds. Forcing $\alpha_\mathrm{T} = 0.6$ may therefore lead to significant error propagation in the characterisation of the non-thermal part of the spectrum.
    \item[-] Radio measurements used for the fit were sparse and broad-band flux densities. The lack of broad continuous sampling of the measurements in sub-bands restricts the diagnostic potential of the analysis. Incorporating the intra-band flux distribution would better constrain the radio spectrum.
    \item[-] Lower frequency measurements (close to 1 GHz and below) do not necessarily display a flux drop that is steep enough to be compliant with the expected cut-off of f-FFA.
\end{enumerate}

Our intention is to revisit the analysis of the radio SED of WR\,147, taking advantage of its high brightness, in order to gain some insight into the interpretation of its behaviour. Given the very long period of the system, we stress that the multi-epoch nature of the data presented in Sect.\,\ref{obs} is not a problem.

\section{Radio data}\label{obs}
\subsection{VLA data}
The main part of our data set was collected with the VLA (Proposal ID: VLA/16A-252), in L-band (1.0 -- 2.0 GHz) and C-band (4.0 -- 8.0 GHz), configured into 8 and 32 spectral windows of 125 MHz width each, in one epoch, on 8-9 August 2016. Each spectral window is subdivided into 64 channels with a width of 2 MHz. At the time of these observations, the array was set in the B configuration, which leads to an angular resolution of 1.01$"$ $\times$ 0.82$"$ for the C--band and 5.20$"$ $\times$ 3.14 $"$ for the L--band. The system is unresolved at this resolution and appears as a point source. 

We processed the data (flagging, calibration, and imaging) using the Common Astronomy Software Applications (\texttt{CASA}) \citep{McMullin2007}. The source used as both bandpass and flux scale calibrator was 3C286.        J2007+404 was used for complex gain calibration.
The bandpass calibrator and flux scale calibrator was observed at the beginning, and the complex gain calibrator was observed just before and after WR\,147. The integration time on WR\,147 was 45 minutes for the L-band and 15 minutes for the C-band. We flagged bad data following the standard procedure for VLA observations. We proceeded with manual flagging to mitigate the impact of Radio Frequency Interferences (RFI), especially in the L-band, which was severely affected. This led to a complete loss of signal in some specific spectral windows. In fine, we obtained 6 valid spectral windows in the L-band, and 24 in C-band. The deconvolution of images was achieved using the {\tt tclean} task in \texttt{CASA}, adopting robust weighting with Briggs parameter set to 0. 

\begin{figure}[h]
\centering
\includegraphics[width=\linewidth]{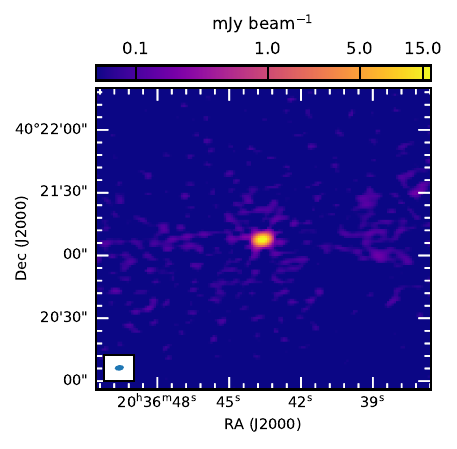}
\includegraphics[width=\linewidth]{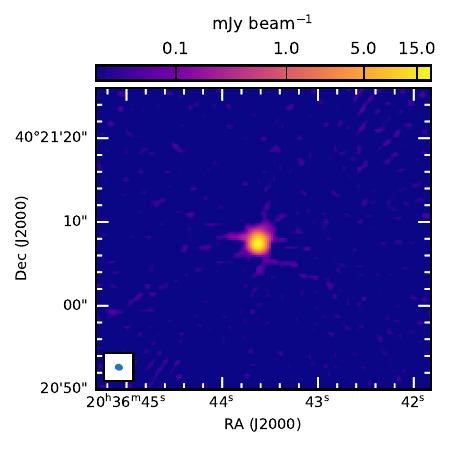}
\caption{Images of field centred about the position of WR\,147 in L-band (upper panel) and C-band (lower panel). The synthesised beam is shown in the bottom left corner.
\label{fig:Images}}
\end{figure}

The flux density of WR\,147 was measured for each spectral window, in both bands. 
Our approach consisted of fitting Gaussians to the point source using the \texttt{imfit} function of \texttt{CASA}. We retrieved the flux density and the statistical error for every image. 
The error bar on the flux density was set to the quadratic sum of the statistical error and a systematic contribution of 2\% of the flux density of the calibrator\footnote{\url{https://science.nrao.edu/facilities/vla/docs/manuals/oss2016A/performance/fdscale}}. The results of all our measurements are summarised in Table\,\ref{tab:flux}. 

\begin{table}
    \centering
        \caption{VLA measurements per spectral window.}
    \label{tab:flux}
    \begin{tabular}{ccccc}
    \hline
        Spw & $\nu$  & $S_\nu$ & $\sigma_{fit}$  & $\sigma$ \\
            & (MHz)  & (mJy) & (mJy) & (mJy) \\
        \hline
         1& 1199 &  23.70 &  0.33 & 0.58\\
         2& 1327 & 24.09  &  0.23 & 0.53 \\
         3& 1455 & 24.93 & 0.23 & 0.55\\
         5& 1711 & 25.81 & 0.24 & 0.57\\
         6& 1839 & 26.52 & 0.23 & 0.58\\
         7& 1967 & 27.17 & 0.30 & 0.62 \\
         \hline
        8 & 4039 & 31.43 & 0.17 & 0.65\\
        9 & 4167 & 31.43 & 0.18 & 0.65\\
        10 & 4295 & 32.17 & 0.18 & 0.67\\
        11 & 4423 & 32.24 & 0.17 & 0.67\\
        13 & 4679 & 30.89 & 0.17 & 0.64\\
        14 & 4807 & 31.88 & 0.17 & 0.66 \\
        16 & 5063 & 31.59 & 0.17 & 0.65\\
        18& 5319 & 32.85 & 0.20 & 0.68\\
        19& 5447 & 32.05 & 0.22 & 0.68 \\
        20& 5575 & 33.50 & 0.27 & 0.72 \\
        21& 5703 & 33.97 & 0.25 & 0.72\\
        22& 5831 & 32.83 & 0.25 & 0.70\\
        24& 6039 & 32.85 & 0.28 & 0.72\\
        26& 6295 & 33.41 & 0.31 & 0.74 \\
        28& 6551 & 32.35 & 0.36 & 0.74 \\
        29& 6679 & 33.48 & 0.32 & 0.74\\
        31& 6935 & 34.50 & 0.37 & 0.78 \\
        32& 7063 & 32.40 & 0.37 & 0.74 \\
        34& 7319 & 34.01 & 0.43 & 0.81 \\
        35& 7447 & 34.11 & 0.41 & 0.80\\
        36& 7575 & 34.16 & 0.40 & 0.79\\
        37& 7703& 32.40 & 0.38 & 0.75\\
        38& 7831 & 33.28 & 0.44 & 0.80 \\
        39& 7959 & 33.01 & 0.38 & 0.76 \\
         \hline
    \end{tabular}
    \tablefoot{The columns express the central frequency, the flux density, the statistical error from the fit $\sigma_\mathrm{fit}$, and the final error on the flux density, for each spectral window. The first section (down to spw\,7) is related to the L-band, and the second one to the C-band.}
\end{table}

\subsection{Previously published data}
For the purpose of our analysis, we used previously published radio measurements of WR\,147. For the higher-frequency part, we used several measurements in the 40 -- 375 GHz range covering a part of the spectrum fully dominated by thermal emission. These data were collected with the VLA and the James Clerk Maxwell Telescope (JCMT). At lower frequencies, we used flux density measurements obtained with the Giant Meterwave Radio Telescope (GMRT). The conservative upper limit at 150\,MHz is coming from the TIFR GMRT Sky Survey (TGSS) \footnote{\url{https://tgssadr.strw.leidenuniv.nl/doku.php?id=start}}, and is taken as three times the typical noise level at the position of the source. Finally, a measurement obtained with the Westerbork Synthesis Radio Telescope (WSRT) has been considered in our discussion. A summary of these other measurements is given in Table\,\ref{otherradio}. As there is no bright radio source in the direct vicinity of our target, the rather poor angular resolution at some of these frequencies is not a problem. 

\begin{table}[h]
\centering
\caption{List of the radio measurements used to complete the SED of WR\,147.}\label{otherradio}
\begin{tabular}{lccc}
\hline
$\nu$ &  Instr. & $S_\nu$ & Date\\
(GHz) &  & (mJy) &  \\
\hline
0.15 & GMRT & $<$ 10 & 2010-2012$^{(a)}$ \\
0.35 & WSRT & 16.0 $\pm$ 4.0 & 1996$^{(b)}$ \\
0.61 & GMRT & 20.6 $\pm$ 0.3 & Aug 2016$^{(c)}$\\ 
\hline
15.0 & VLA & 49.3 $\pm$ 2.0 & Jan 1985$^{(d)}$ \\
22.5 & VLA & 59.1 $\pm$ 2.0 & Jan 1985$^{(d)}$ \\
42.0 & VLA & 83.0 $\pm$ 2.0 & Apr 1995$^{(e)}$ \\
231 & JCMT & 280 $\pm$ 30 & Mar 1991$^{(f)}$ \\
273 & JCMT & 292 $\pm$ 15 & Mar 1991$^{(f)}$ \\ 
375 & JCMT & 357 $\pm$ 70 & Mar 1991$^{(f)}$ \\ 
\hline
\end{tabular}
\tablefoot{$a$: \citet{Intema2017}; $b$: \citet{setia}; $c$ \citet{Benaglia2020}; $d$: \citet{Churchwell1992}; $e$: \citet{Contreras1996}; $f$: \citet{williamswr147}.}
\end{table}

\section{Spectral analysis}\label{spec}
It is important to note that the data presented in Tables\,\ref{tab:flux} and \ref{otherradio} show a positive trend with increasing frequency. This behaviour is, as will be confirmed by our spectral analysis, attributable to the exceptionally high thermal brightness of the WR wind. Whatever part of the composite radio SED that is considered for WR\,147, the spectral index is positive. This is in contrast to the usual expectation of a negative (or flat) index, at least in some parts of the spectrum, for synchrotron sources. 

The spectral analysis is based on Python scripts, making use of Matplotlib \citep{Hunter2007}, Pandas \citep{McKinney2010}, SciPy \citep{Virtanen2020} and NumPy \citep{Harris2020} packages.

\subsection{Foreground-FFA}\label{fFFA}
To achieve a preliminary view of the parameters required to investigate the SED, we first used the higher frequency data only to characterise the part of the radio spectrum dominated by thermal emission (see the lower part of Table\,\ref{otherradio}). We used the following equation:
\begin{equation}\label{eqT}
   S_\nu^\mathrm{T} = A\,\nu^{\alpha_\mathrm{T}} 
\end{equation}
\noindent that corresponds to the first term of Eq.\,\ref{eq:SEDfFFA}. This was fitted to the flux density measurements using a weighted least-squares method. The best-fit parameters obtained were $A = 8.40 \pm 0.65$ (in units of mJy\,GHz$^{-\alpha_\mathrm{T}}$) and $\alpha_\mathrm{T} = 0.62 \pm 0.02$. The uncertainties represent the standard errors derived from the covariance matrix. We obtained a reduced $\chi^2$ equal to 2.15. We caution that the thermal index determined at this stage of the data analysis should not be over-interpreted as the data points used so far are certainly contaminated by some non-thermal emission arising from the colliding-wind region.

\begin{figure}[h]
\centering
\includegraphics[width=\linewidth]{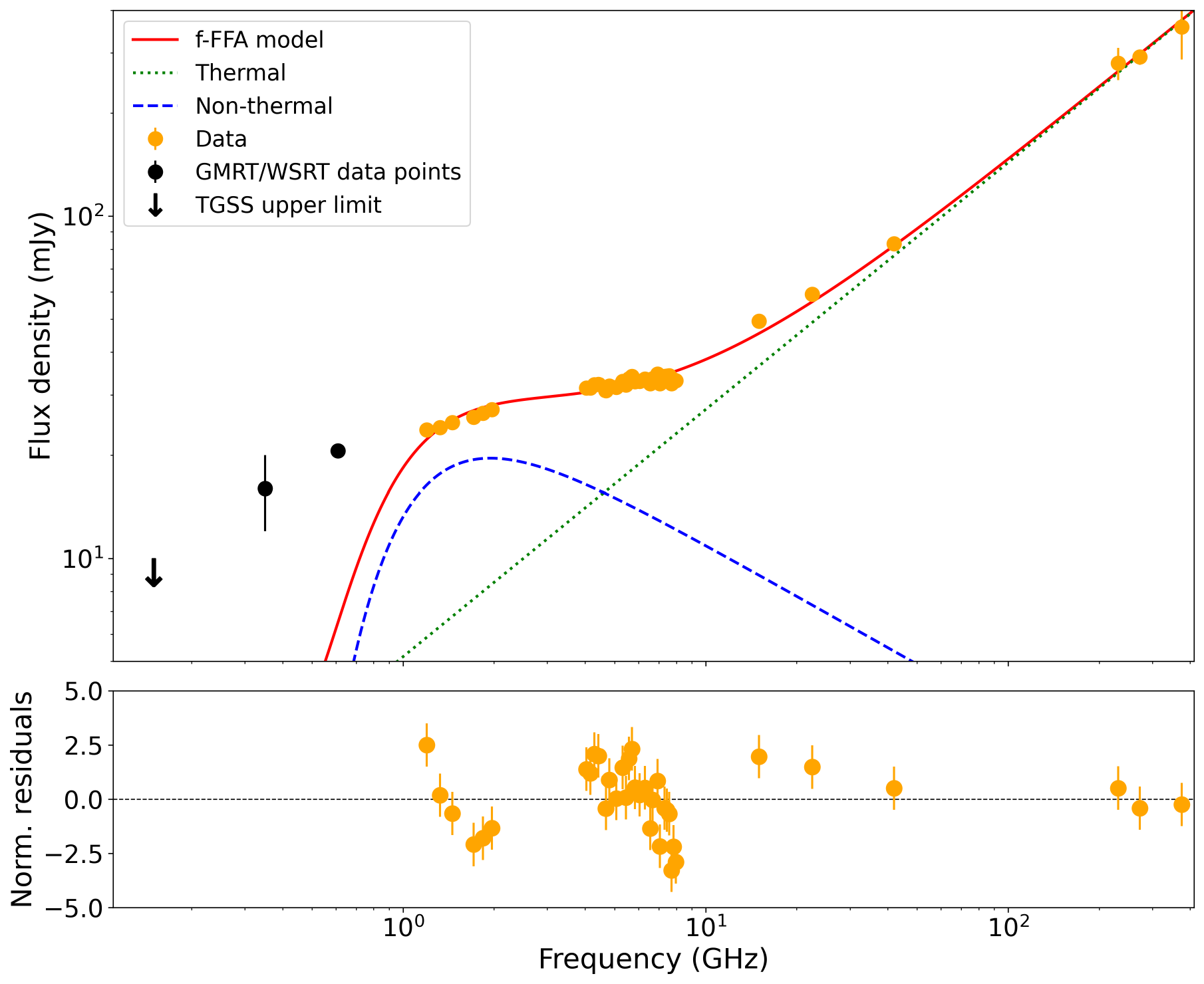}
\caption{Upper panel: radio SED of WR\,147 fitted with Model C. The GMRT and WSRT measurements, not considered for the fit, are over-plotted. Lower panel: normalised residuals in the sense data minus the model.
\label{fig:PlotfFFA}}
\end{figure}

As a next step, we used the high-frequency previous measurements combined with our VLA data quoted in Table\,\ref{tab:flux} to fit the composite radio spectrum defined by Eq.\,\ref{eq:SEDfFFA}. We first fixed the values of $A$ and $\alpha_\mathrm{T}$ derived above based on Eq.\,\ref{eqT}, allowing only $B$ and $\alpha_\mathrm{NT}$ to vary (Model A). The best-fit parameters are reported in Table\,\ref{tab:fFFA}. We then allowed the two frozen parameters to vary as well and obtained a statistically consistent fit (Model B). This led to a significant steepening of the thermal power-law component. However, we note that the non-thermal spectral index is shallower than expected for a synchrotron source, assuming relativistic electrons accelerated by DSA in high Mach number adiabatic shocks (for monoatomic gas). This makes this result not fully satisfactory. We then explored potential alternative solutions, forcing $\alpha_\mathrm{NT}$ to be 0.5, which is the canonical value for the shock conditions specified above (Model C). 
The fit reaches a statistical quality similar to that of Model B, but with apparently more relevant physical values. The best-fit with Model C is shown in Fig.\,\ref{fig:PlotfFFA}. The f-FFA turnover occurs at a frequency such that the optical depth is 1, i.e. 0.98\,GHz.

We finally added the 0.61\,GHz GMRT data point to the analysis. We were unable to obtain a relevant fit. The $\chi^2$ reaches values greater than 100 (for 31 $dof$), and the parameters adopt physically irrelevant values. In particular, the non-thermal index is almost flat, which does not make any sense for a pure optically thin synchrotron spectrum. To illustrate the difficulty of reconciling the GMRT data point with the other measurements above 1\,GHz, we over-plotted the 610\,MHz flux density on the best-fit result shown in Fig.\,\ref{fig:PlotfFFA}. One clearly sees that the GMRT measurement is not compliant with the expected exponential drop in the synchrotron flux at low frequencies due to f-FFA. This is in agreement with the lack of a physically relevant fit when including the GMRT data, as reported above. The over-plotted WSRT measurement is affected by a quite large relative error. For the same reason as explained for the GMRT data point, no adequate fit including the WSRT measurement could be obtained with the model described in this section. These results motivated us to explore an alternative modelling approach that is more compatible with the trend displayed by our L-band data and extends down to the GMRT measurement frequency (and to some extent the WSRT data point, despite its large relative error). In addition, the extrapolation of the trend displayed by the 610\,MHz and L-band measurements is clearly above the conservative upper limit at 150 MHz. This suggests that another turnover occurs at quite low frequencies. This will be discussed in Sect.\,\ref{secondturnover}.

\begin{table}[h]
\centering
\caption{Best-fit parameters for the f-FFA SED of WR\,147 (without considering the GMRT and WSRT data points).}\label{tab:fFFA}
\begin{tabular}{lccc}
\hline
 &  {\it Model A}& {\it Model B} & {\it Model C}\\
 \hline
$A$ & 8.40$^\dagger$  & 4.04 $\pm$ 0.73 & 5.16 $\pm$ 0.30 \\
$\alpha_\mathrm{T}$ & 0.62$^\dagger$  & 0.77 $\pm$ 0.04 & 0.72 $\pm$ 0.02  \\
$B$ & 40.97 $\pm$ 3.60 & 31.40 $\pm$ 2.11 & 34.70 $\pm$ 1.50  \\
$\alpha_\mathrm{NT}$ &  0.95 $\pm$ 0.05 & 0.34 $\pm$ 0.09 & 0.50$^\dagger$ \\
$\nu_\mathrm{f-FFA}$ & 1.20 $\pm$ 0.06 & 0.86 $\pm$ 0.08 & 0.98 $\pm$ 0.03 \\
$\chi^2/dof$ & 121.21/33 & 75.66/31 & 91.58/32\\
\hline
\end{tabular}
\tablefoot{$dof$ stands for degree of freedom. $A$ and $B$ are expressed in units of mJy\,GHz$^{-\alpha_\mathrm{T}}$ and mJy\,GHz$^{-\alpha_\mathrm{NT}}$, respectively. $\nu_\mathrm{f-FFA}$ is expressed in GHz. $\dagger$ Fixed values.}
\end{table}

\subsection{Internal-FFA}\label{iFFA}

\subsubsection{Context of the model}\label{iFFAcontext}
\begin{figure}[h]
\centering
\includegraphics[width=\linewidth]{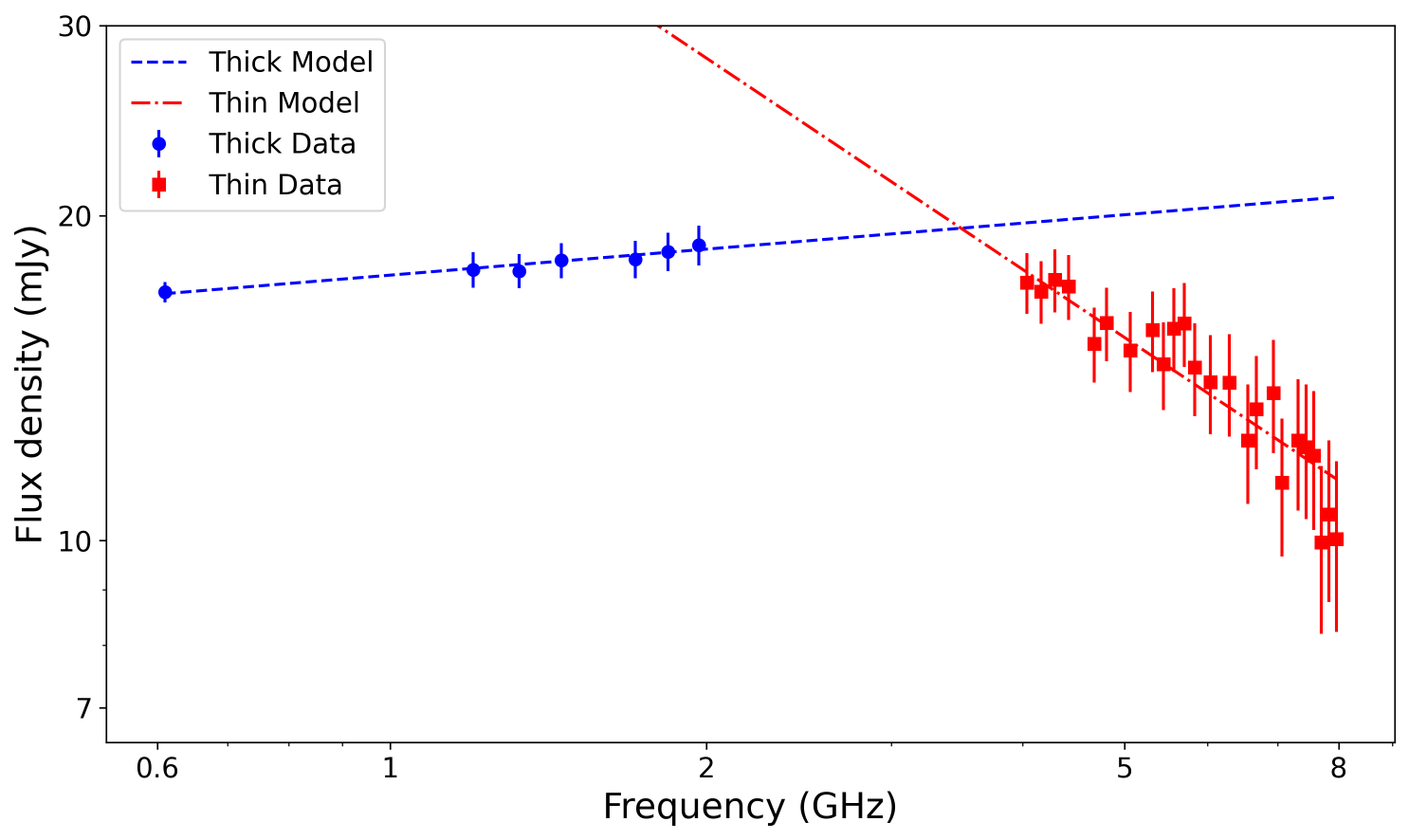}
\caption{GMRT, L-band and C-band data (corrected for the thermal contribution) with their best-fit power-law models. \label{fig:plotNTPL}}
\end{figure}

To emphasise the lack of a low-frequency exponential drop and the presence of an apparent monotonous trend below the turnover (which seems to be located at a few GHz), we first focused on the non-thermal emission part of the spectrum. To do so, we used the parameters $A$ and $\alpha_\mathrm{T}$ from Model C (see Sect.\,\ref{fFFA}) as an estimator of the thermal emission component. We subtracted the thermal model values from the measured flux densities with appropriate error propagation to focus on the non-thermal contribution of the spectrum. We then separately fitted power-law models to lower frequency (GMRT + L-band) and higher frequency (C-band) residual data. We will refer to these regimes as optically thick and optically thin, respectively. The two power-law functions are defined as follows:
\begin{align}
S_\mathrm{Thin}^\mathrm{NT} &= B_\mathrm{Thin} \nu^{\alpha_\mathrm{Thin}}, \\
S_\mathrm{Thick}^\mathrm{NT} &= B_\mathrm{Thick} \nu^{-\alpha_\mathrm{Thick}}.
\end{align}
For the lower frequency data, we obtain $B_\mathrm{Thick} = 17.29 \pm 0.40$ and $\alpha_\mathrm{Thick} = 0.08 \pm 0.05$ ($\chi^2/dof = 0.25/5$). This suggests a slightly positive slope that is compatible with an almost flat trend. For the higher frequency part, the power-law fit yields $B_\mathrm{Thin} = 40.79 \pm 14.20$ and $\alpha_\mathrm{Thin} = 0.62 \pm 0.21$ ($\chi^2/dof = 1.47/22$). This spectral index value is, within the error bars, compatible with the optically thin synchrotron index obtained with Model C. The results are displayed in Fig.\,\ref{fig:plotNTPL}, where all relevant data points (corrected for thermal emission) and models are plotted. One sees that all data plotted in that figure are compatible with a switch between two power-law regimes, with a turnover occurring between 3 and 4 GHz. 

This behaviour is reminiscent of the occurrence of internal free--free absorption (hereafter, i-FFA). This has been addressed, for instance, by \citet{deBruyn1976}, in the framework of the study of synchrotron-emitting galaxies where no significant foreground absorption is expected and where no low-frequency exponential drop is measured. The spatial coincidence between a radio emission region and the presence of electrons responsible for absorption is well known and documented, for instance, when dealing with optically thick free--free emission or optically thick synchrotron emission. In these two examples, the electron populations responsible for the absorption are of thermal and relativistic nature, respectively. However, one should not disregard the scenario where the synchrotron emission arises from a region that is at least partly filled with a thermal plasma where FFA is (even weakly) active within it, without any further strong f-FFA. A PACWB such as WR\,147 is a good candidate to comply with this scenario. The synchrotron emission region is measured with an offset of about 0.6$''$ from the position of the WN star. At a distance of 1.7\,kpc, this translates into a projected separation of about 1.5$\,\times\,10^{16}$\,cm ($\sim$ 1000\,AU). At such a distance, even the strong WN stellar wind is not very dense, drastically reducing the strength of f-FFA. However, the coexistence between thermal and relativistic electrons in the vicinity of the colliding-wind region still has to be considered. Even though the post-shock plasma temperature in between the two standing shocks is too high to warrant significant FFA, one should keep in mind that relativistic electrons are not confined in this inter-shock region. Their spatial extension also coincides with some cooler plasma, upstream of both shocks, which is in principle capable of significantly contributing to FFA. 

Let us set the equation of radiative transfer in the following form,
\begin{equation}\label{eq:radtransf}
I_\nu = \frac{J_\nu}{4\pi\,Q_\nu}\,\bigg[1 - \exp(-\tau_\nu)\bigg],
\end{equation}
\noindent where $J_\nu$ is the synchrotron emissivity per unit volume and $Q_\nu$ is an equivalent modified FFA coefficient averaged across the source region. In the exponential factor, one identifies the optical depth defined as $\tau_\nu = \int\,Q_\nu\,ds$, where the integration is calculated over the distance element $ds$. This emphasises the scaling of the optical depth with the FFA absorption coefficient.

The dependence on the frequency of the two key factors in Eq.\,\ref{eq:radtransf} are expressed as follows: $J_\nu \propto \nu^{-\alpha_\mathrm{NT}}$ and $Q_\nu \propto \nu^{-\beta}$, where $\beta$ is a parameter that depends significantly on the distribution of the absorbing material in the region of interest. The dependence on the density distribution of the spectral index of the spectrum where i-FFA is important has been discussed, for instance, by \citet{deBruyn1976} for some specific geometries. The aforementioned study also considers the influence of the spatial distribution in relativistic electrons and that of the local magnetic field on the slope of the internally absorbed spectrum. Thus, a wide range of measured slopes is expected, depending on the interplay between the distributions of all these parameters. Addressing these spatial distributions in detail would require in-depth magnetohydrodynamic and radiative modelling that is out of the scope of the present study.

At high frequencies, the dependence of the specific intensity of the source on the frequency is reduced to that of $J_\nu$. This is the usual optically thin synchrotron regime. At low frequencies, $Q_\nu$ is high enough to allow the exponential factor of Eq.\,\ref{eq:radtransf} to vanish. The frequency dependence appears therefore as follows,
\begin{equation}\label{eq:iFFAthick}
I_\nu = \frac{J_\nu}{4\pi\,Q_\nu} \propto \nu^{-\alpha_\mathrm{NT} + \beta}\, ,
\end{equation}
\noindent which defines the optically thick regime in the occurrence of i-FFA.

\subsubsection{Confrontation to the data}\label{iFFAfit}
We will consider a model accounting for i-FFA in the radio spectrum of WR\,147 defined as follows,
\begin{equation}\label{eq:SEDiFFA}
S_\nu = A\,\nu^{\alpha_\mathrm{T}} + B'\,\nu^{-\alpha_\mathrm{NT} + \beta}\,\left[1 - e^{-(\nu/\nu_\mathrm{i-FFA})^{-\beta}}\right],
\end{equation}
\noindent where $B'$ is a factor that basically scales with the ratio of the synchrotron emissivity to the amplitude of the FFA, and $\nu_\mathrm{i-FFA}$ is the i-FFA turnover frequency where the optical depth is equal to one. In the framework of this phenomenological model, $\beta$ is typically a parameter that has to be tuned to allow the model to reproduce the radio spectrum where i-FFA is important. Other parameters have been defined when introducing Eq.\,\ref{eq:SEDfFFA} or in Sect.\,\ref{iFFAcontext}.

\begin{figure}[h]
\centering
\includegraphics[width=\linewidth]{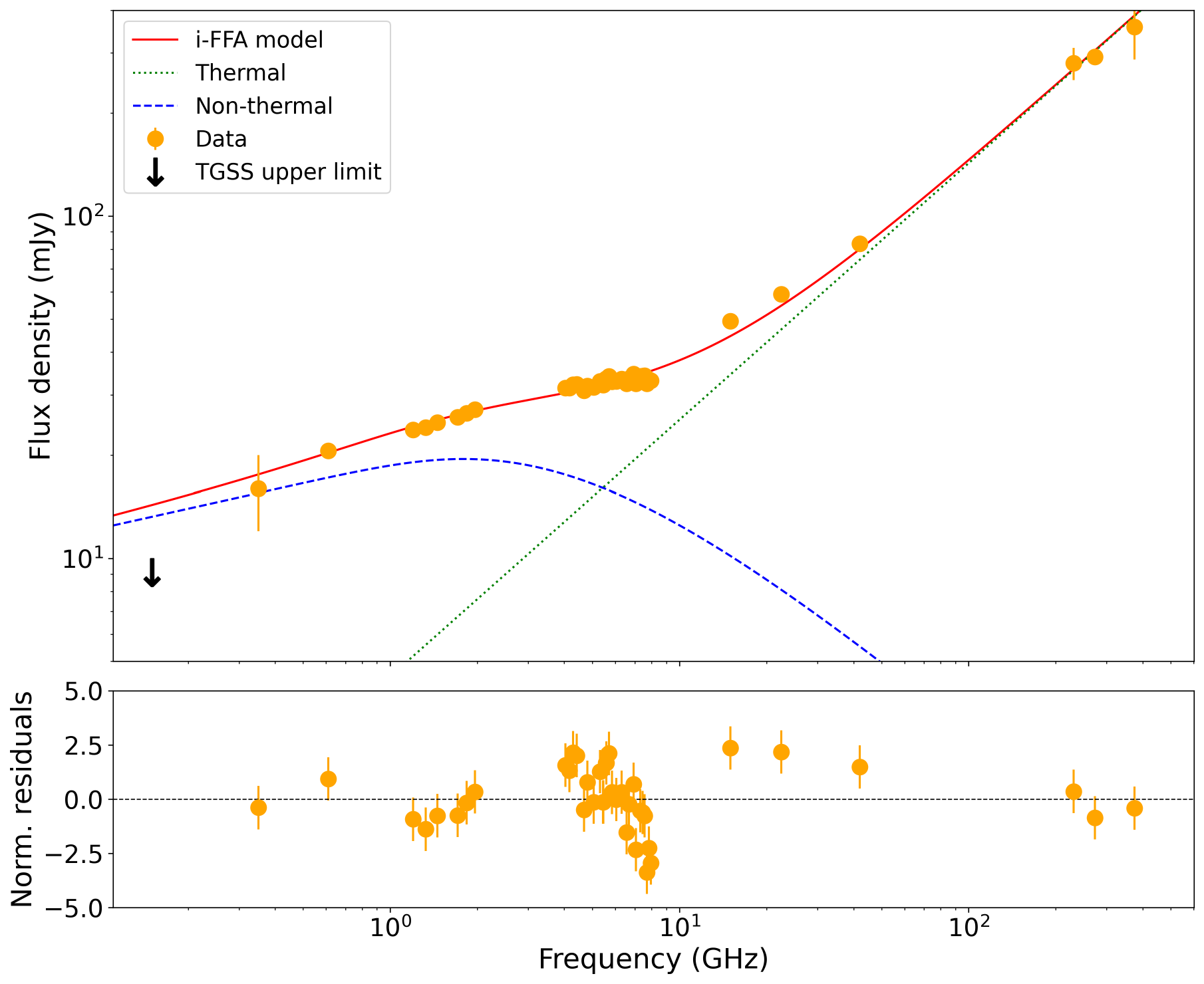}
\caption{Upper panel: best-fit i-FFA model and data, with $\alpha_\mathrm{T}$ and $\alpha_\mathrm{NT}$ fixed to 0.75 and 0.7, respectively. The TGSS upper limit is shown with a black arrow. Lower panel: normalised residuals in the sense data minus model. \label{fig:PlotiFFAResiduals}}
\end{figure}

We fitted this model to the entire series of data quoted in Table\,\ref{tab:flux} and Table\,\ref{otherradio} (not counting the TGSS upper limit). The convergence of the fitting procedure is highly dependent on the spectral indices, and the uncertainties on these parameters prevent us from reaching a valid solution in a straightforward way. We therefore decided to explore a parameter grid by freezing $\alpha_\mathrm{T}$ and $\alpha_\mathrm{NT}$, allowing other parameters ($A$, $B'$, $\nu_\mathrm{i-FFA}$ and $\beta$) to vary. The parameter grid is presented in Table\,\ref{tab:GridiFFA}. Our results deserve a few comments:
\begin{enumerate}
\item[-] Our best-fit results with this model are much better than any f-FFA model including the 610\,MHz GMRT and the 350\,MHz WSRT measurements  (see Sect.\,\ref{fFFA}). Given its large relative error, the WSRT data point doesn't provide any strong constraint on the fit: removing it from the data series leads to the same results within error bars. 
\item[-] The statistically best results are obtained for rather steep thermal and optically thin non-thermal spectra. However, this trend reaches its limit when $\alpha_\mathrm{T} = 0.8$ is assumed, for increasing values of $\alpha_\mathrm{NT}$. No adequate fit could be obtained in this parameter space region. This corresponds indeed to a situation where the model is not able to achieve a valid solution, as both thermal and non-thermal components concur to carve a cavity in the SED that is not compliant with the measured slope seen in the data.
\item[-] Among our model grid, the statistically best result corresponds to $\alpha_\mathrm{T} = 0.75$ and $\alpha_\mathrm{NT} = 0.7$. This case is plotted in Fig.\,\ref{fig:PlotiFFAResiduals}. Some slight statistical improvement is obtained for steeper non-thermal indices, but the physical relevance of these values is questionable.
\item[-] The rather steep thermal spectral index is in agreement with several measurements of thermal spectra from WR stars \citep{Nugis1998,Benaglia2019}.
\item[-] The steep non-thermal index points to a significant deviation with respect to the expectation of relativistic electrons accelerated through linear DSA, by high Mach number adiabatic shocks with monoatomic gas. This situation has already been encountered when measuring the index of other PACWBs. We mention, for example, the cases of HD\,167971 \citep{SanchezBermudez2019,DeBecker2024}, HD\,168112 \citep{Blomme2005,Blomme2024}, or Apep \citep{Marcote2021}. In case of highly efficient particle acceleration, this may be attributed to shock modification that leads to a steepening of the lower energy part of the relativistic spectrum. Another potential explanation comes from the relative motion of scattering centres in the fluid, leading to a modified effective compression for back-scattering of relativistic electrons resulting in a steeper particle spectrum \citep{Pittard2021}. We clarify that our approach explicitly assumes a uniform non-thermal index across the full synchrotron spectrum. This may not be appropriate when the effect of inverse Compton (IC) cooling is pronounced. The net effect is a steepening of the relativistic electron distribution at high Lorentz factor values, resulting in a steepening of the optically thin synchrotron index as well \citep{Pit2}. This effect is directly proportional to the local radiative energy density in the colliding-wind region, with target photons for IC scattering coming from the stellar photospheres. Therefore, the very long separation in WR\,147 is not in favour of a strong IC cooling effect, especially in the spectral range where the synchrotron emission from WR\,147 is measured.
\item[-] We caution that the thermal emission component is likely to behave in a more complex way as assumed here in our modelling. It is possible that the simple power-law thermal model may not perfectly reflect the measurements at higher frequencies. It has indeed been shown by \citet{Nugis1998} that the thermal emission may display a non-uniform index between the GHz and the THz regimes. In addition, our results may be affected by some variability in the wind emission as suggested by \citet{Churchwell1992} and \citet{Contreras1996}. Inaccuracies in the modelling of the thermal emission component are affecting our view of the non-thermal emission as well, therefore limiting our capability to model the radio SED. 
\end{enumerate}

\begin{table*}[h]
\centering
\caption{Model grid used in the framework of i-FFA with best-fit parameters.}
\label{tab:GridiFFA}
\begin{tabular}{ccccccc}
\hline
$\alpha_\mathrm{T}$ & $\alpha_\mathrm{NT}$ & $A$ & $B'$ & $\nu_\mathrm{i-FFA}$ & $\beta$ & $\chi^2$ \\
  &  & (mJy\,GHz$^{-\alpha_\mathrm{T}}$) & (mJy\,GHz$^{-\alpha_\mathrm{NT}}$) & (GHz) &  &  \\
\hline
0.7 & 0.5 & $5.39 \pm 0.13$ & $21.05 \pm 0.84$ & $2.14 \pm 0.39$ & $0.89 \pm 0.07$ & 97.62 \\
0.7 & 0.7 & $5.52 \pm 0.12$ & $18.24 \pm 0.29$ & $4.12 \pm 0.62$ & $0.91 \pm 0.04$ & 84.91 \\
0.75 & 0.5 & $4.38 \pm 0.11$ & $20.72 \pm 0.54$ & $3.07 \pm 0.57$ & $0.81 \pm 0.05$ & 86.99 \\
0.75 & 0.7 & $4.52 \pm 0.10$ & $18.90 \pm 0.24$ & $5.20 \pm 0.70$ & $0.89 \pm 0.03$ & 76.07 \\
\hline
\end{tabular}
\tablefoot{The number of $dof$ in each case is 34 (38 data points for 4 free parameters).}
\end{table*}

\begin{figure}[h]
\centering
\includegraphics[width=\linewidth]{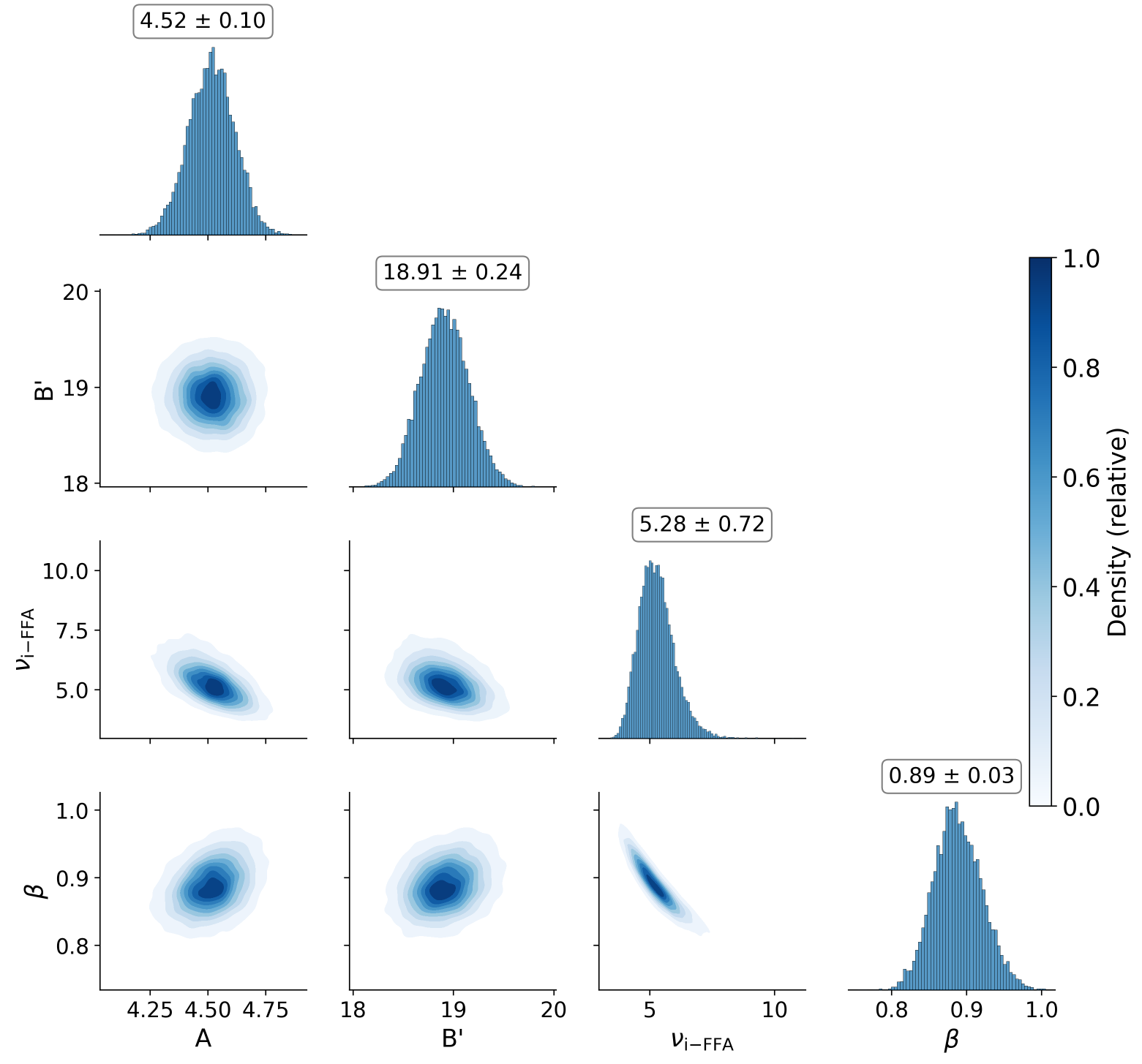}
\caption{Corner plot resulting from the Monte-Carlo simulations. The best-fit parameters with 1-$\sigma$ uncertainty are displayed on top of each histogram. \label{fig:mc}}
\end{figure}

As a complementary approach, we proceeded with a last fit attempt using Monte-Carlo error propagation on the non-linear fit. We considered the case where spectral indices are fixed such as $\alpha_\mathrm{T} = 0.75$ and $\alpha_\mathrm{NT} = 0.7$. The results are fully consistent with those displayed in Table\,\ref{tab:GridiFFA}. We produced a corner plot (showing the posterior-like distributions and covariances) using the Seaborn visualisation package \citep{Waskom2021} that is presented in Fig.\,\ref{fig:mc}.

The diagonal panels display the marginal distributions of each model parameter, derived from 10000 Monte Carlo realisations of the fit. These histograms reflect the spread and central tendency of the parameter estimates. The correlation diagrams display a significant anti-correlation between $\beta$ and $\nu_\mathrm{i-FFA}$: high $\beta$ values correlate with an i-FFA turnover at lower frequencies. These two parameters are crucial to reproduce the non-thermal component. For a general discussion on the dependence of the SED on various parameters, we refer to our Sect.\,\ref{disc} where a slightly different parametrisation will be adopted for the sake of clarity of the interpretation.

\subsection{Other turnover scenarios}\label{otherTO}

Apart from FFA, one should also investigate the relevance of considering other turnover processes likely to affect a synchrotron spectrum at lower frequencies.

\subsubsection{Synchrotron Self-Absorption (SSA)}\label{SSA}
A classical case of SSA in a homogeneous emitter would allow us to expect an optically thick synchrotron spectrum with a spectral index equal to 2.5, which is much steeper than that displayed in the low-frequency part of Fig.\,\ref{fig:plotNTPL}. Although flatter optically thick spectra may to some extent be explained by the superposition of many self-absorbed contributions peaking at various frequencies, it could also be explained by a non-uniform source \citep[see e.g.][]{deBruyn1976}. However, for SSA to be active, a strong requirement is that the number density in relativistic electrons must be high enough to allow these absorbers to dominate the solution of the radiative transfer equation up to the frequency where the turnover is measured, i.e. above 1\,GHz according to our results on WR\,147.

The SSA turnover frequency ($\nu_\mathrm{SSA}$) can be expressed as a function of the flux density at the turnover frequency ($S_\mathrm{SSA}$) and of the angular size of the source. This translates into a measurement of the compactness required for a synchrotron source to produce significant SSA at a given frequency. Based on the equation used by \citet{wr147setia}, we can express the angular size $\theta_\mathrm{SSA}$ (in mas),

\begin{equation}
\theta_\mathrm{SSA} \approx 0.08\,S_\mathrm{SSA}^{1/2}\,B_\mathrm{CWR}^{1/4}r\,\nu_\mathrm{SSA}^{-5/4},
\end{equation}
\noindent where $S_\mathrm{SSA}$ is expressed in mJy, $B_\mathrm{CWR}$ is the magnetic field strength in the colliding-wind region expressed in mG and $\nu_\mathrm{SSA}$ is expressed in GHz. According to our measurements, the turnover is occurring close to 3 GHz where the non-thermal flux density is about 15 mJy. Assuming a magnetic field strength of the order of 5 mG as determined by \citet{wr147setia}, we obtain an angular size that allows SSA to affect the spectrum of about 0.1\,mas. This value is about three orders of magnitude smaller than the source extension measured by high angular resolution observations \citep{williamswr147}. The synchrotron source, as measured in WR\,147, is therefore not compact enough for SSA to be responsible for the turnover.

\subsubsection{Razin-Tsytovitch (RT)}\label{RT}
This plasma effect consists of the inhibition of the synchrotron emission process in a magnetised thermal plasma. In practice, it leads to a low-frequency exponential drop of the synchrotron flux at a turnover frequency (in Hz) expressed as follows \citep{Pacholczyk1970}: 
\begin{equation}\label{eq:freqRT}
    \nu_\mathrm{RT} = 20 \frac{n_\mathrm{e}}{B_\mathrm{CWR}},
\end{equation}
\noindent with $n_\mathrm{e}$ being the number density of thermal electrons and $B_\mathrm{CWR}$ the magnetic field (both expressed in cgs units) in the synchrotron emission region. 

The pre-shock gas density ($n$) of the WR wind at the location of the synchrotron emission region can be calculated using this relation:
\begin{equation}
n = \frac{{\dot M}}{4\,\pi\,r^2\,\mu\,m_\mathrm{H}\,V_\infty},
\end{equation}
\noindent where ${\dot M}$ is the mass loss rate, $V_\infty$ is the terminal velocity (expressed in cgs units). According to \citet{Hamann2019}, these parameters are $10^{-3.8}$\,M$_\odot$\,yr$^{-1}$ and 1000\,km\,s$^{-1}$, respectively. For the mean molecular weight, we adopt a value of 3.21 \citep{WR147parameters}. For the distance $r$, we assumed the projected linear separation corresponding to the angular offset 0.6$''$ at the distance of 1.7\,kpc, i.e. 1.5\,$\times$\,10$^{16}$\,cm. As the inclination of the system is probably greater than 0, this constitutes a lower limit on the actual linear separation (leading thus to an upper limit on $n$). We obtain a gas number density of about 6600\,cm$^{-3}$. Let us assume a density that is typical of the post-shock region. Once again, this leads to a conservative limit, as the full synchrotron emission region is not limited to the post-shock region. With a gas compression ratio of 4, valid for adiabatic high Mach number shocks (for monoatomic gas), the post-shock density is about 26500\,cm$^{-3}$. Since the mean number of electrons per ion is of the order of 1, $n$ constitutes a valid estimate of $n_\mathrm{e}$. For orders of magnitude considerations, we assume $B_\mathrm{CWR}$ = 1\,mG. Since the RT turnover frequency is higher for weaker magnetic fields, this constitutes a conservative limit as compared to the 5\,mG value proposed by \cite{wr147setia}. As a result, we estimate that the RT cut-off frequency is about 500\,MHz only. This value, which is likely an upper limit given the assumptions made to calculate it, is much lower than the turnover frequency inferred from our SED fitting. 

Despite the large uncertainties on several parameters that influence the value of $\nu_\mathrm{RT}$, it is clear that its value is much lower than the turnover measured in the radio data. Therefore, the influence of the Razin-Tsytovitch effect on the radio spectral energy distribution of WR~147 can therefore be ruled out.

\subsubsection{The second turnover}\label{secondturnover}
The upper limit at 150\,MHz indicates that a regime change occurs at a frequency of a few 100\,MHz. The data coverage at our disposal shows that two turnovers are needed to interpret the non-thermal component of the radio SED of WR\,147. In addition to the turnover that we attribute to internal FFA at a few GHz, a second turnover is required to explain the lower emission level at 150\,MHz. Unfortunately, the large relative error in the WSRT measurement at 350\,MHz does not provide any strong constraint on the position of this second turnover.  Regarding the discussion above, one may consider two potential scenarios for the second turnover:
\begin{enumerate}
\item[-] {f-FFA:} The lack of strong f-FFA at a few GHz is explained by the low column of material coming from the wind, given the large separation between the WR and the colliding-wind region. However, this lower density material has the potential to produce f-FFA at lower frequencies (as the f-FFA turnover frequency scales with the square of the density). Thus, the full SED of WR\,147 may display both FFA regimes, foreground and internal, at two very different frequencies.
\item[-] {Razin-Tsytovitch:} According to our discussion in Sect.\,\ref{RT}, a comfortable upper limit for $\nu_\mathrm{RT}$ is several 100\,MHz. This admits the possibility that this process may be occurring somewhere between 150 and 610 MHz.
\end{enumerate}
A key factor is the local magnetic field strength: a value significantly greater than 1\,mG (as assumed in Sect.\,\ref{RT}) would certainly rule completely out Razin-Tsytovitch. It is thus tempting to favour f-FFA to explain the occurrence of the lower frequency turnover. However, whatever the interpretation one could envisage, the data clearly display a double-turnover behaviour, which is reported for the first time to the best of our knowledge for this class of objects.

\section{Discussion}\label{disc}

\subsection{Comparison to previous detailed modelling}\label{compmodel}
It is certainly worth commenting on the SED presented in this work in light of the modelling of the radio emission from WR + O systems of the kind of WR\,147 presented in previous seminal works. \citet{Doug} proposed results from high-resolution hydrodynamical simulations and solutions to the radiative transfer equation, leading to predictions of synchrotron radio spectra. FFA is thus accounted for consistently across a model grid. Among their results, they have shown that a change in the orientation of the system leads to changes in the radio spectrum, including some configurations where the FFA turnover is much shallower than a pure exponential cut-off. This happens typically for inclinations lower than 60$^\circ$, while close to 90$^\circ$ (line-of-sight along the orbital plane) a pure f-FFA behaviour is seen as a consequence of strong FFA by the WR wind. This is quite complementary to what is presented in this paper, where we suggest that a low inclination favours the measurement of i-FFA. We caution, however, that the spectra presented by \citet{Doug} are not fully and directly comparable to the case of WR\,147. Although the WC + O case they considered is highly reminiscent of WR\,147, their assumed separation values are much shorter than the one we report on in this paper.

A similar behaviour is shown in synthetic spectra published by \citet{Pit2} who extended the modelling by \citet{Doug} considering the temporal and spatial evolution of relativistic electron populations. Similar curves for various inclinations show a shallower low-frequency drop for inclinations that do not favour strong f-FFA. We also note that the sampling of the radio emission from WR\,147 as a function of frequency that was available at that time was not sufficient in the lower frequency part to provide strong constraints on the modelling. An updated application of their modelling based on the data set presented in this paper, taking into account the recent revision of several important parameters (mass-loss rate, distance...) would certainly be instructive, and relevant for a comparison to the phenomenological interpretation framework we propose in this work.

\subsection{Toward a generic radio SED for PACWBs}\label{sedpacwb}
Our results show that the usual a priori model of foreground FFA of the synchrotron radio emission from PACWBs is not necessarily the most appropriate in every case. Our analysis of the radio SED of WR\,147 shows that the internal FFA constitutes a relevant alternative, more consistent with the lower frequency measurements used in this work. However, f-FFA should not be rejected as a major source of synchrotron emission attenuation, as usually claimed for several PACWBs \citep{Doug,Doug140,DeBecker2019,Benaglia2020,Saha2023,Blanco2024}. Therefore, it is relevant to clarify the circumstances favouring the measurement of the signature of f-FFA or i-FFF in colliding-wind massive binaries. The requirement for f-FFA to be prominent is the existence of a foreground slab of ionised material, with a density that is high enough to warrant a high free--free optical depth. In such a scenario, the synchrotron source is in the background of the absorbing material, and its radio flux undergoes a significant, if not substantial, exponential drop. Alternatively, i-FFA relies on the idea that a thermal plasma distributed in the synchrotron emitting region contributes to FFA, but the treatment of the radiative transfer leads rather to a shallower attenuation at lower frequencies. It should be noted that the impact of i-FFA is rather moderate compared to that of f-FFA and produces a change of slope below the turnover frequency. \\

The discussion of these two FFA conceptions in their appropriate astrophysical context shows that they are not mutually exclusive. A colliding-wind region in a PACWB may be active at synchrotron emission, potentially modulated by some i-FFA, before being more severely absorbed at lower frequencies through f-FFA. As both FFA formalisms occur at lower frequencies, it is straightforward to anticipate that a strong FFA by a dense foreground environment will prevent us from identifying the signature of i-FFA in the system. This idea can be formalised by setting a generic SED function defined as follows:
\begin{equation}\label{eq:generic}
S_\nu = C_1\,\nu^{\alpha_\mathrm{T}} + \frac{C_2}{C_3}\,\nu^{-\alpha_\mathrm{NT} + \beta}\,\bigg[1 - e^{-C_3\,\nu^{-\beta}}\bigg]\,e^{-(\nu/\nu_\mathrm{f-FFA})^{-2.1}},
\end{equation}
\noindent where the first term represents the thermal emission from the stellar winds and the second term stands for the synchrotron emission component affected by i-FFA before being further modulated by the exponential f-FFA turnover. $C_1$ is the normalisation parameter for the thermal emission component. For a more general discussion, we adopted a slightly different parametrisation as compared to previous sections for i-FFA: $C_2$ scales with the synchrotron emissivity and $C_3$ with the amplitude of the absorption coefficient, while $\beta$ represents the index for the frequency dependence of the i-FFA coefficient. We explicitly represent the normalisation parameter for the non-thermal term as scaling with the ratio of synchrotron emissivity and i-FFA absorption coefficients (see Eq.\,\ref{eq:radtransf}). This is especially important as $C_3$ is also a scaling factor of the i-FFA turnover frequency. In other words, considering various i-FFA turnover locations also has an impact on the normalisation of the non-thermal term. This representation allows us to discuss further the interplay between key parameters. We clarify that Eq.\,\ref{eq:generic} assumes that $\nu_\mathrm{SSA}$ and $\nu_\mathrm{RT}$ are significantly lower than the turnover frequencies for f-FFA and i-FFA. In other words, this description is valid, provided neither SSA nor RT participate in the modulation of the measured radio emission. This condition is very likely fulfilled for most, if not all, PACWBs where FFA dominates the low-frequency spectrum. On the one hand, SSA requires the source to be compact enough to allow self-absorption to dominate. On the other hand, the RT turnover frequency increases with the density, whereas that of FFA scales with the square of the density, strengthening the idea that any higher density region is more prone to be mainly affected by FFA. We also note that the plasma in the colliding-wind region may contribute to some thermal emission, especially at higher frequencies \citep{Pittard2010}. However, this may be significant provided the medium is dense enough, and this occurs typically in quite short period systems that are not compliant with the usual profile of known PACWBs. For this reason, we do not explicitly consider such a thermal contribution here. In principle, this should be embedded in the first term, and this term is most probably dominated by the thermal emission from the winds themselves (in particular, when dealing with WR winds).

To explore the parameter space relevant to Eq.\,\ref{eq:generic}, we started with parameter values close to those obtained for WR\,147. We checked for the influence of all the parameters on the generic SED by varying these values one by one. The parameter grid is detailed in Table\,\ref{tab:genericgrid}. 

\begin{table*}[h]
\centering
\caption{Parameter grid used to investigate the behaviour of the generic SED defined by Eq.\,\ref{eq:generic}.}
\label{tab:genericgrid}
\begin{tabular}{lccccc}
\hline
{\it Set} & $C_1$ & $C_3$ & $\nu_\mathrm{f-FFA}$ & $\alpha_\mathrm{NT}$ & $\beta$ \\
  & (mJy\,GHz$^{-\alpha_\mathrm{T}}$) & (GHz$^{\beta}$) & (GHz) &  &  \\
\hline
(a) & 5.0 & 5.0 & [0.2; 0.8; 1.4]  & 0.7 & 0.85 \\
(b) & [1.0; 2.0; 5.0] & 5.0 & 0.2  & 0.7 & 0.85 \\
(c) & 5.0 & 5.0 & 0.2 & 0.7 & [0.65; 0.85; 1.05] \\
(d) & 5.0 & [2.0; 5.0; 8.0] & 0.2 & 0.7 & 0.85 \\
(e) & 5.0 & 5.0 & 0.2 & [0.5; 0.7; 0.9] & 0.85 \\
\hline
\end{tabular}
\tablefoot{Two parameters were fixed: $C_2 = 90.0$ and $\alpha_\mathrm{T} = 0.7$.}
\end{table*}

\begin{figure}[h]
\centering
\includegraphics[width=\linewidth]{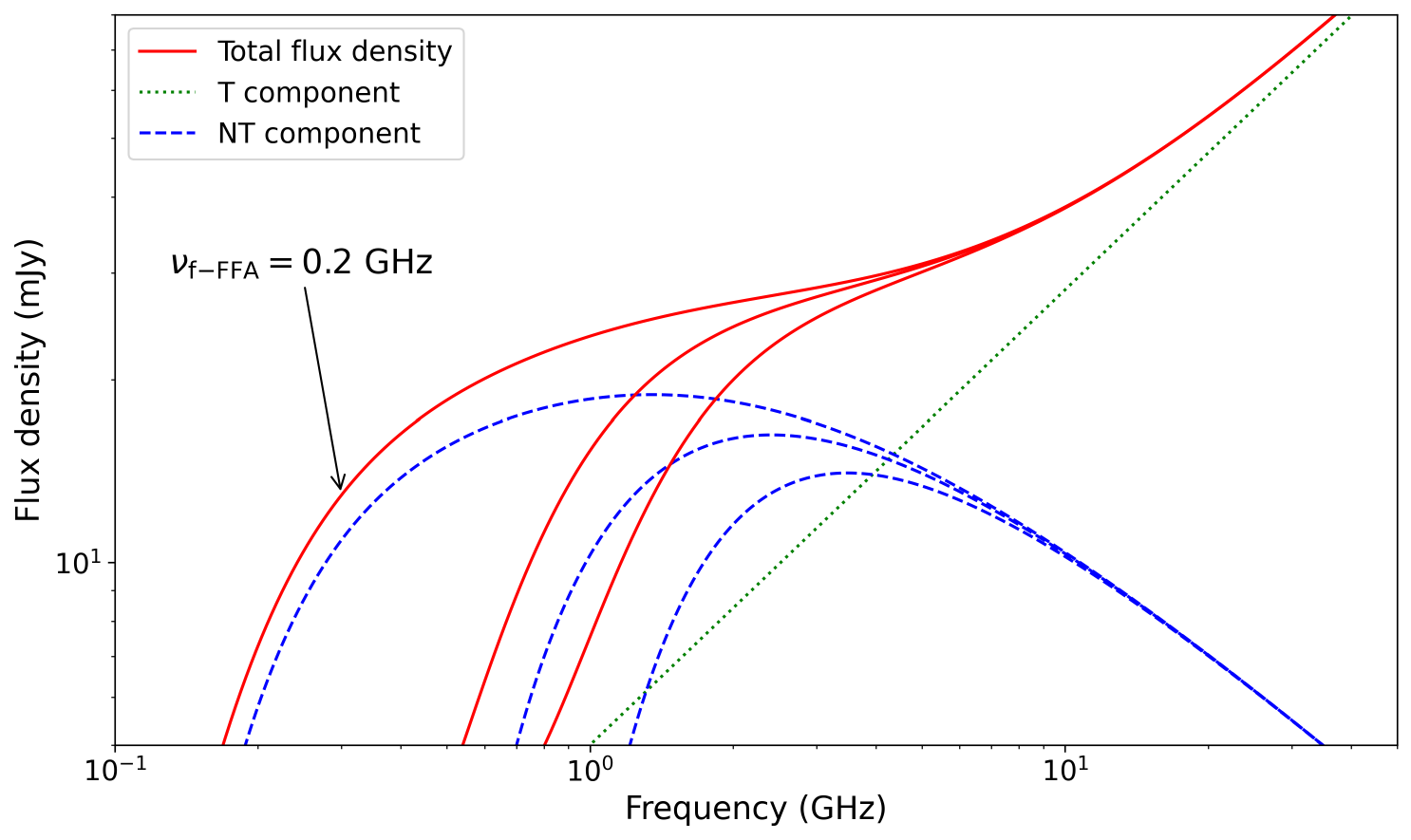}
\caption{Generic SED of a PACWB displaying the effect of both i-FFA and f-FFA. Three different values for $\nu_\mathrm{f-FFA}$ are considered (Set (a)), with the curve extending further to the left standing for the lower $\nu_\mathrm{f-FFA}$ considered. The thermal (dotted line) and non-thermal (dashed line) components are individually displayed. \label{fig:generic_a}}
\end{figure}

We first investigated the influence of f-FFA on an internally free--free absorbed SED such as that of WR\,147 (Set (a)). The results are shown in Fig.\,\ref{fig:generic_a}. We adopted a set of values for $\nu_\mathrm{f-FFA}$ equal to 0.2\,GHz, 0.8\,GHz and 1.4\,GHz, respectively. The lower limit of the set of values considered is low enough to warrant a significant part of the SED to display the signature of i-FFA. With the parameters considered for Set (a), the f-FFA turnover frequency of 1.4 GHz still allows identifying the signature of i-FFA, since $\nu_\mathrm{i-FFA}$ is greater than $\nu_\mathrm{f-FFA}$. However, one can anticipate that a very strong f-FFA characterised by a higher turnover frequency would completely suppress the non-thermal component. The relative positions of both foreground and internal turnovers are critical for the detectability of synchrotron sources among PACWBs (see Sect.\,\ref{Incidence} for a discussion).

\begin{figure*}[h]
\centering
\includegraphics[width=0.48\linewidth]{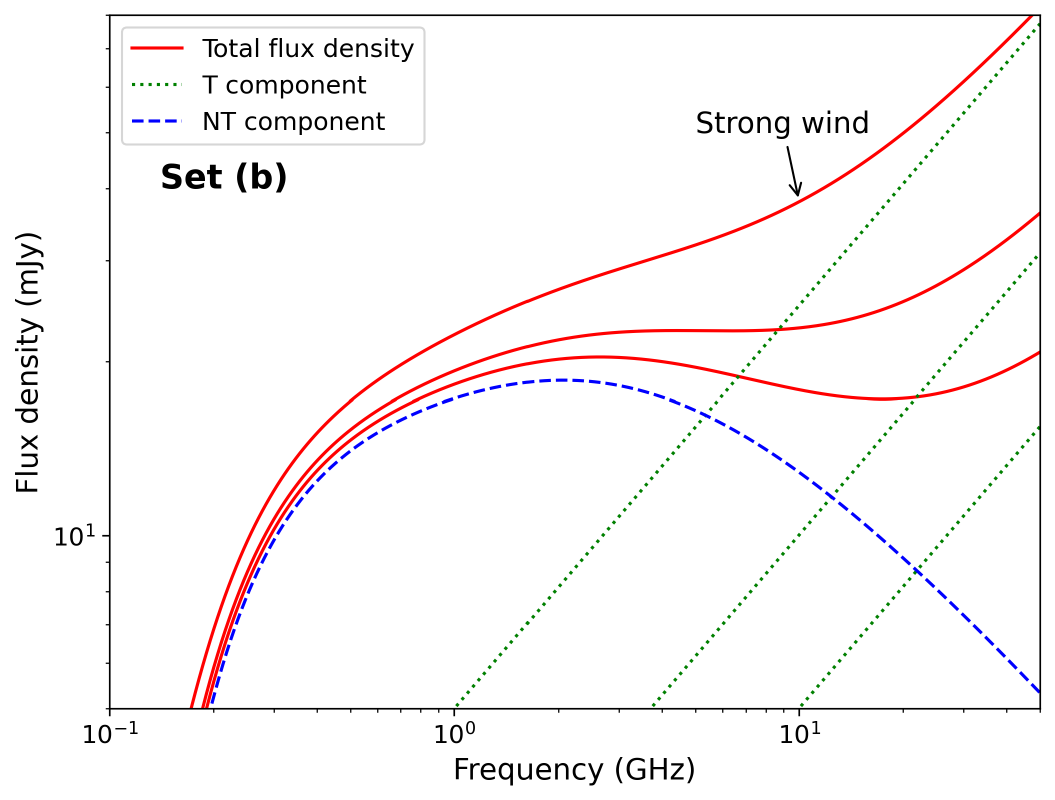}
\includegraphics[width=0.48\linewidth]{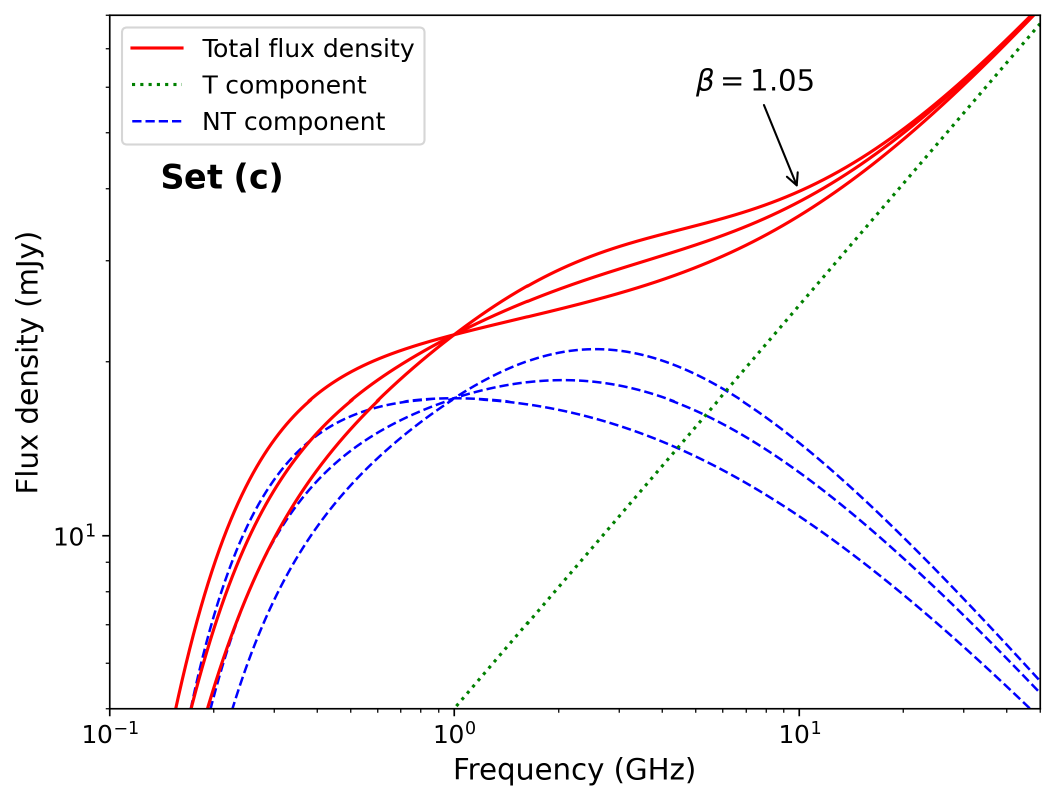} \\
\includegraphics[width=0.48\linewidth]{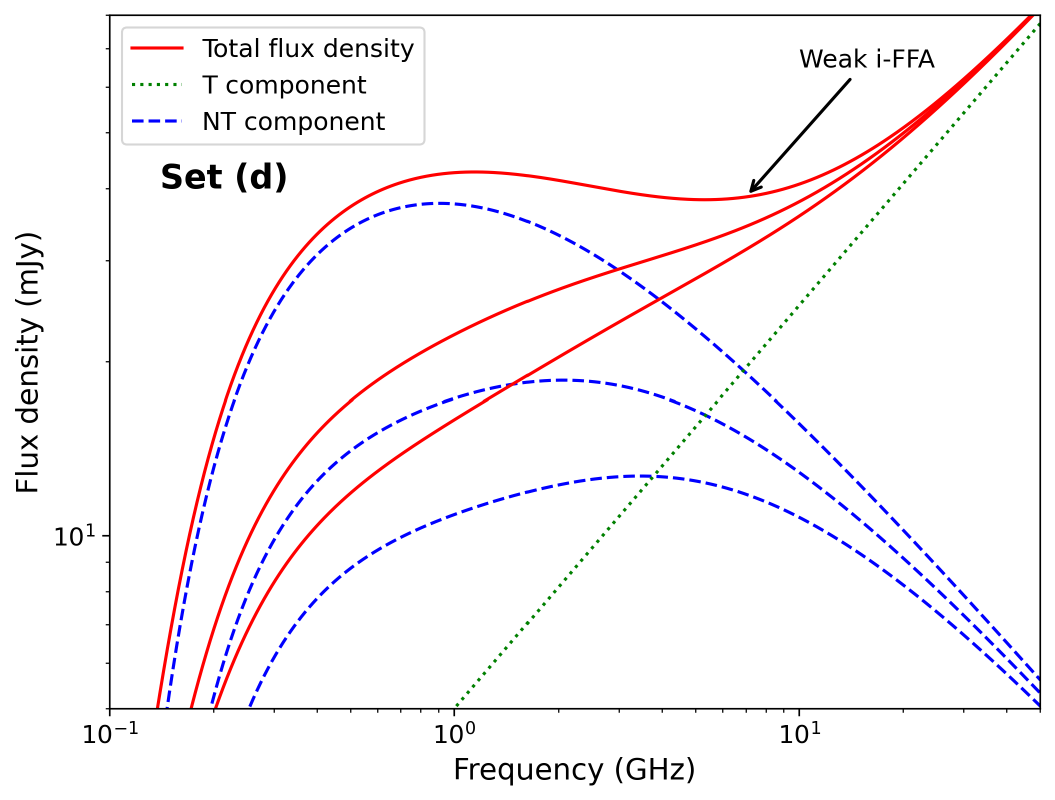}
\includegraphics[width=0.48\linewidth]{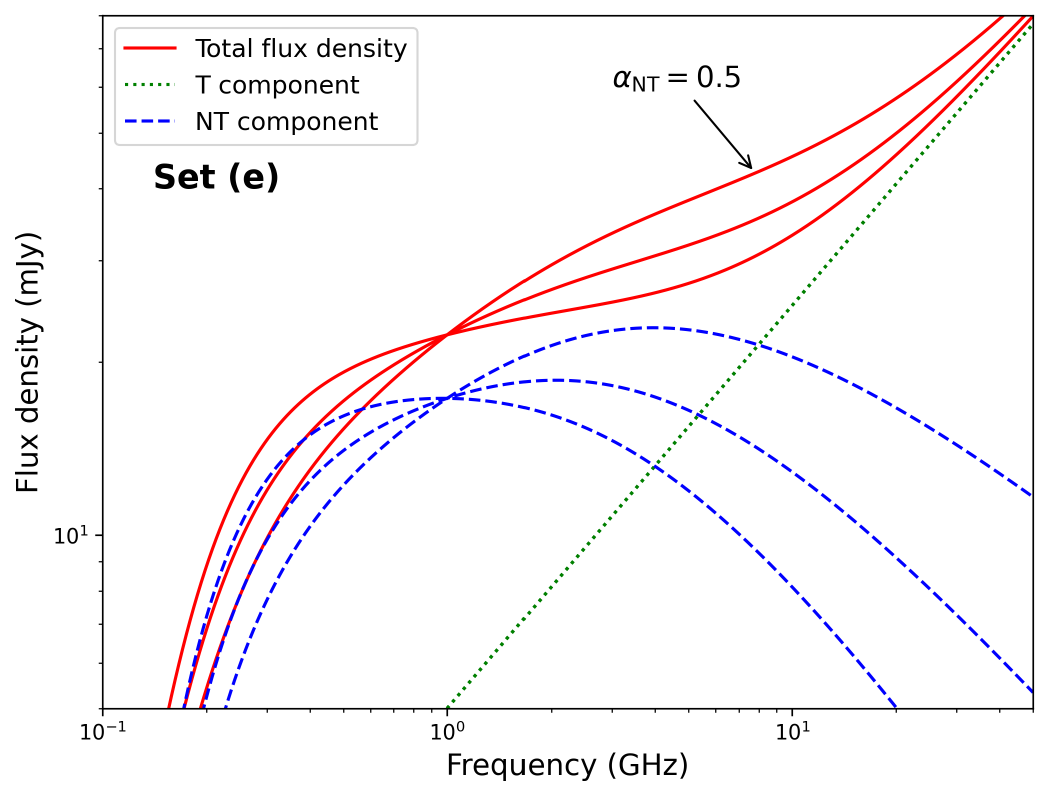}
\caption{Generic SED of a PACWB displaying the effect of both i-FFA and f-FFA, for Sets (b), (c), (d) and (e) (see Table\,\ref{tab:genericgrid}). For details on the meaning of the various curves, see text. \label{fig:generic_bcde}}
\end{figure*}

In the case of Set (b), we varied the amplitude of the thermal emission component ($C_1$), all other parameters being set to the values of Set (a). We adopted the minimum value for $\nu_\mathrm{f-FFA}$ to prevent f-FFA from hiding the effect of other parameters on the non-thermal component of the SED. The results are displayed in the upper left panel of Fig.\,\ref{fig:generic_bcde}. We did not consider a $C_1$ value greater than the typical one we obtained for WR\,147 (pointed as the 'Strong wind' case in the figure). The WN wind is already expected to be among the brightest thermal emitters. However, lower values are worth considering. The net effect of lowering the thermal emission component is an enhancement in the appearance of the synchrotron component, allowing some parts of the SED to be characterised by a flat, or even negative index (more compliant with the usual expectation for a synchrotron source). The contrast between the non-thermal and thermal components could in principle be further enhanced for not too dense O-type winds. However, one should also keep in mind that much weaker winds would inject less kinetic power into the shocks in the colliding-wind region, likely leading to a lowering of the synchrotron luminosity as well.

The effect of the variation of the i-FFA index ($\beta$) on the SED is illustrated in the upper right part of Fig.\,\ref{fig:generic_bcde} (Set (c)). As emphasised in Eq.\,\ref{eq:generic}, this parameter influences both $\nu_\mathrm{i-FFA}$ and the measured index in the part of the SED displaying the signature of i-FFA. The steeper value of this index is highlighted in the SED plot. It corresponds to the curve that peaks at the highest flux density above 1\,GHz while it drops more steeply than others below that frequency. As the resulting spectral index below $\nu_\mathrm{i-FFA}$ is $-\alpha_\mathrm{NT} + \beta$, a higher $\beta$ value steepens the positive slope in the optically thick part. In contrast, the shallower $\beta$ value allows the non-thermal component to reach a broader maximum that is interrupted by f-FFA. The interplay between the two effects (i.e. steepening/flattening of the optically thin synchrotron region and flattening/steepening of the optically thick part), leads to a balance point close to 1 GHz (see Appendix\,\ref{app_A}). It is interesting to note that this frequency corresponds to a spectral position where the flux density is not dependent on the i-FFA index. Changing the values of the other parameters leads only to negligible changes in the spectral position of this balance point. 

The investigation of the influence of the strength of i-FFA requires some caution and is presented in Set (d). It is clear from Eq.\,\ref{eq:iFFAthick} that the normalisation factor for the non-thermal component is basically made of the ratio between the synchrotron emissivity and the modified i-FFA absorption coefficient, the latter being also present in the exponential accounting for i-FFA. The results are shown in the lower left part of Fig.\,\ref{fig:generic_bcde} where the weakest i-FFA case (corresponding to $C_3$ = 2.0) is annotated for the sake of clarity, with increasing strength (thus, increasing $C_3$ values) successively for lower curves. The weakest i-FFA case allows the optically thin synchrotron spectrum to prevail over a wider frequency range, down to a lower $\nu_\mathrm{i-FFA}$ that is approaching $\nu_\mathrm{f-FFA}$. In contrast, the strongest i-FFA case presents two turnovers: a smoother one for i-FFA and a steeper one at a lower frequency for f-FFA. This shows that, in the latter configuration, i-FFA can initiate a drop in the synchrotron flux, which is much lower in amplitude than that due to f-FFA.

Finally, we show the effect of the optically thin synchrotron spectral index ($\alpha_\mathrm{NT}$) on the SED (Set (e)) in the lower right part of Fig.\,\ref{fig:generic_bcde}. On top of the initial value of 0.7, we assumed a value of 0.5 (the canonical index) and a much steeper value (0.9). A spectral index as steep as the latter value is suggested by some measurements and this is still a matter of debate. A high steepness may arise in principle in cases of significant shock modification leading to a steeper electron distribution in the lower energy part of the relativistic particle spectrum \citep[see e.g.][]{DeBecker2024}, or be the result of the relative motion of the scattering centres in the flow \citep{Pittard2021}. In addition to the change in the optically thin part of the spectrum, we also notice an alteration of the measured index below $\nu_\mathrm{i-FFA}$. The general behaviour is similar to the one depicted when measuring the impact of $\beta$ in Set (c), including the occurrence of a balance point close to 1 GHz. Here again, both the optically thin and optically thick parts of the synchrotron spectrum are affected with opposite trends when $\alpha_\mathrm{NT}$ varies: the steepening of the optically thin index leads to a flattening of the optically thick one, and vice versa. This results in a skewing of the non-thermal component of the radio spectrum. In this case as well, in the parameter space explored to produce these generic plots, the position of the balance point is not altered by changing values of the other parameters.

\subsection{Incidence on the detection of PACWBs}\label{Incidence}
According to the broadened picture of the radio SED of PACWBs discussed in this work, sources can be categorised following the relative positions of both (f-FFA and i-FFA) turnover frequencies:
\begin{enumerate}
\item[-] Case I: $\nu_\mathrm{f-FFA} \ll \nu_\mathrm{i-FFA}$. This is the most favourable situation for the detection of the non-thermal component. Even below the i-FFA turnover, there is a spectral range where the source flux is not heavily attenuated, enhancing the probability of identifying its PACWB nature. However, the non-thermal component is fainter than that of the intrinsic optically thin emission, leading to a significant underestimation of the synchrotron emission power.
\item[-] Case II: $\nu_\mathrm{f-FFA} \sim \nu_\mathrm{i-FFA}$. This intermediate case is characterised by a severe drop in the synchrotron flux at a frequency that is high enough to significantly affect the detectability of synchrotron emitters. The low-frequency synchrotron drop is exponential, with no highly pronounced intermediate regime where the i-FFA signature can be measured. The synchrotron emission above $\nu_\mathrm{i-FFA}$ is certainly representative of the optically thin regime.
\item[-] Case III: $\nu_\mathrm{f-FFA} \gg \nu_\mathrm{i-FFA}$. This extreme case of FFA is likely to fully suppress the non-thermal component from the radio SED. Such cases prevent us from identifying the PACWB nature of the system.
\end{enumerate}

According to the analysis presented in Sect.\,\ref{iFFA}, the example of WR\,147 is clearly compatible with case I. The spectral extension of the i-FFA-dominated region can only be determined by good-sensitivity measurements down to very low frequencies. A fundamental requirement for case I is a lack of dense absorbing medium surrounding the synchrotron emission region, which can typically be achieved provided the orbital period is long enough with an orbit not seen under a too high inclination angle. The systems more compliant with case II are a priori among the most abundant. The reason is that in most known PACWBs, the orbital period is short enough to allow a large amount of stellar wind material surrounding the synchrotron emission region to produce significant f-FFA. The emblematic PACWB WR\,140, with its well-documented radio emission \citep{Doug140}, belongs to this category. It displays indeed a bright non-thermal radio emission that is severely attenuated close to periastron, when the non-thermal emission region is deeply embedded in the dense WR wind. One may also mention the example of WR\,98a, not detected at 1.4\,GHz in GMRT data, but known as a particle accelerator thanks to measurements at several GHz \citep{Blanco2024}. Alternatively, WR\,11 constitutes the perfect example of a case III system. Its radio SED displays a perfect thermal behaviour down to a few MHz \citep{Benaglia2019}, even though it is known as a particle accelerator thanks to $\gamma$-ray observations \citep{MartiDevesa2020}.

An important dimension of the behaviour of PACWBs worth considering is their variability. Especially in an eccentric system, the density of the colliding-wind material varies as a function of the orbital phase. In addition to changing the conditions for intrinsic synchrotron emission itself, these changes will impact both internal and foreground absorption processes. The higher density encountered close to the periastron will favour cases II and III, while approaching the apastron may push the system to leave case III if the period is long enough, potentially favouring the detection of the non-thermal radio emission (case II). This is reminiscent of the case of WR\,133, which displays a purely thermal emission spectrum with the exception of a clear detection of non-thermal emission at another epoch \citep{DeBecker2019}. In the context of the scenario described here, WR\,133 is potentially switching between case II and case III depending on the orbital phase.

\section{Conclusions}\label{conc}
We investigated the radio SED of the very-long period PACWB WR\,147. Our analysis was based on L-band and C-band VLA data with a quality high enough to allow measuring the flux density of the source in 6 and 24 spectral windows, respectively, in both bands. These intra-band measurements were combined with others at higher and lower frequencies to discuss the SED of WR\,147 from metric to millimetric wavelengths. Independently of any modelling, the measured SED displays positive indices across the full radio spectrum, in contrast with the expectation of synchrotron sources. This is fully explained by the exceptionally bright thermal emission from the dense WN wind that is not spatially resolved from the synchrotron emission region in our measurements.

We fitted a theoretical SED made of a thermal (stellar wind emission) and a non-thermal (synchrotron emission) component. We first accounted for FFA using the usual foreground absorption approach (f-FFA), which turns out to be inadequate to reproduce the lower-frequency measurements. Alternatively, we proposed that FFA due to a population of thermal electrons distributed in the synchrotron emission region (i-FFA) is more compatible with the shallower trend measured in the data, which is in contrast to the expected exponential drop of f-FFA. In this model, in addition to its dependence on the abundance of absorbers in the emission region, the i-FFA absorption coefficient is characterised by an index ($\beta$) that is related to the distribution of material. Establishing the relation between $\beta$ and, notably, the distribution of the electron density for the specific geometry of a colliding-wind region is out of the scope of this paper. However, in the framework of our simple parametric modelling, $\beta$ is a parameter that has to be tuned to allow the model to reproduce the radio spectrum where i-FFA is important.

Although the simple phenomenological model does not perfectly reproduce the data, it is qualitatively in good agreement with the general trends observed in the data set. The occurrence of i-FFA could be explained by the very long period of the system. The winds collide at a very long separation, preventing the unshocked wind material surrounding the synchrotron emission region from being dense enough to contribute to severe f-FFA at the frequencies of available data. This could also further indicate that the orbital plane is seen under a rather low inclination angle. A much shorter separation with a different inclination would certainly have led to a f-FFA turnover at much higher frequencies, therefore hiding the signature of i-FFA. However, we stress that the upper limit at 150\,MHz clearly points to the presence of a second turnover at a few 100 MHz. Even though we cannot fully reject a Razin-Tsytovitch origin for this second turnover, it may indicate the occurrence of f-FFA at very low frequencies. WR\,147 is the only system displaying this double-turnover behaviour to the best of our knowledge.\\

The considerations developed in the framework of the example of WR\,147 motivated us to discuss a more generic phenomenological SED for PACWBs. We propose that, in the absence of a full hydro-radiative model, the qualitative SED of PACWB could be made of two terms: (i) a term for thermal emission (accounting for the emission from the stellar winds) and (ii) a non-thermal term, where modulation by both i-FFA and f-FFA is taken into account (see Eq.\,\ref{eq:generic}). This generic SED is appropriate for quantifying the relative importance of the thermal and non-thermal components, along with the relative positions of the i-FFA and f-FFA turnover frequencies. The latter point has a huge effect on the detectability of the non-thermal radio component. While a system with $\nu_\mathrm{f-FFA} << \nu_\mathrm{i-FFA}$ allows for the detection of a non-thermal excess across a potentially wide range of the spectrum, a case where $\nu_\mathrm{f-FFA} >> \nu_\mathrm{i-FFA}$ leads more likely to a substantial suppression of the synchrotron emission, preventing its identification as a PACWB. The intermediate case is certainly more typical of most known PACWBs identified through their synchrotron radio emission. The time-variability of the radio emission from PACWBs is also worth being discussed in line with the terms introduced in our work. 

The phenomenological description of the SED of PACWBs proposed in this paper is meant to provide a quick look at the radio emission from these systems. The simple interpretation framework it offers has the benefit of not being restricted to considering a canonical synchrotron source behind a screen of absorbing material. However, for a more detailed description, a full magnetohydrodynamic modelling including radiative effects is required. This is especially important when addressing the details of the distribution of thermal and non-thermal electrons, along with the magnetic field, across the source of interest.

\begin{acknowledgements}
The authors would like to thank the referee for a fair and positive report including constructive suggestions that helped to improve the discussion. This research is part of the PANTERA-Stars collaboration, an initiative aimed at fostering research activities on the topic of particle acceleration associated with stellar sources\footnote{\url{https://www.astro.uliege.be/~debecker/pantera}}. This research has made use of NASA's Astrophysics Data System Bibliographic Services. The National Radio Astronomy Observatory is a facility of the National Science Foundation operated under cooperative agreement by Associated Universities, Inc. MT acknowledges financial support from the STAR Research Unit (ULiège) for a first year PhD grant. This publication benefits from the support of the Wallonia-Brussels Federation (Belgium) in the context of the FRIA Doctoral Grant awarded to ABB.
\end{acknowledgements}

\bibliographystyle{aa} 
\bibliography{bibfile} 

\begin{appendix}
\twocolumn
\section{Complementary discussion on the dependence of the radio SED on spectral indices}\label{app_A}

\begin{figure}[h]
\centering
\includegraphics[width=0.95\linewidth]{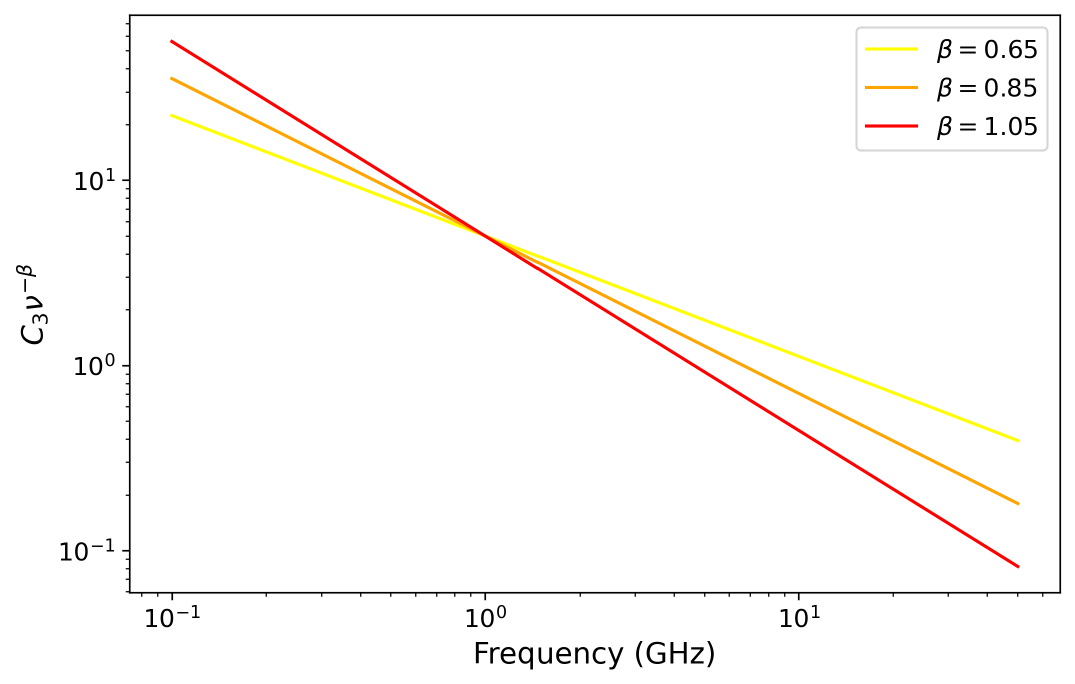}
\includegraphics[width=0.95\linewidth]{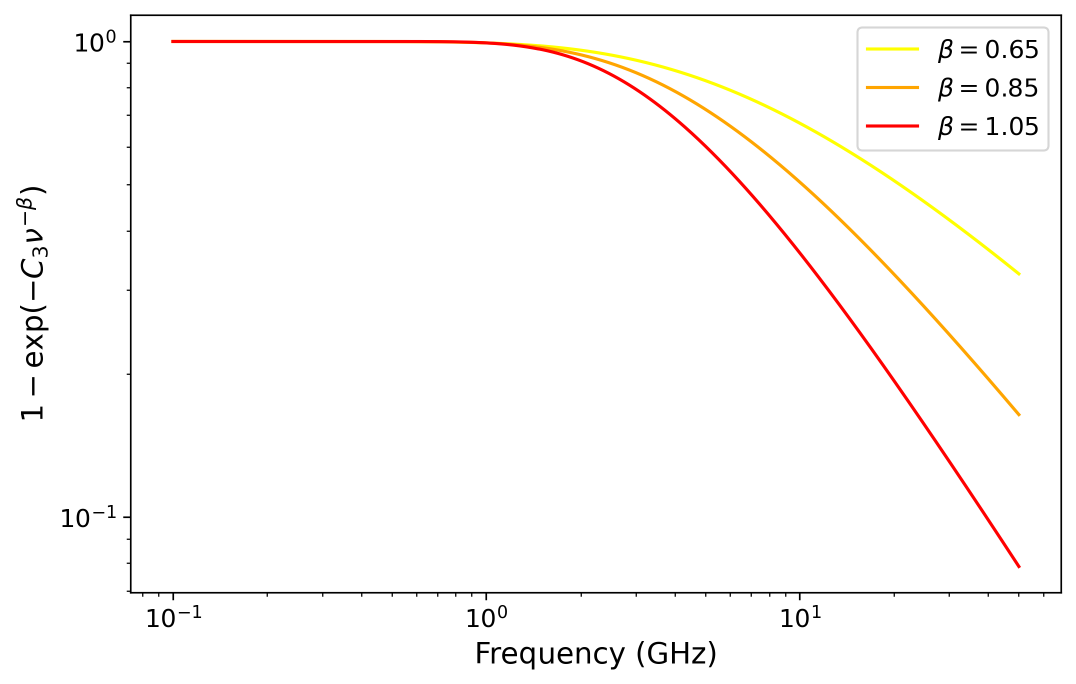}
\includegraphics[width=0.95\linewidth]{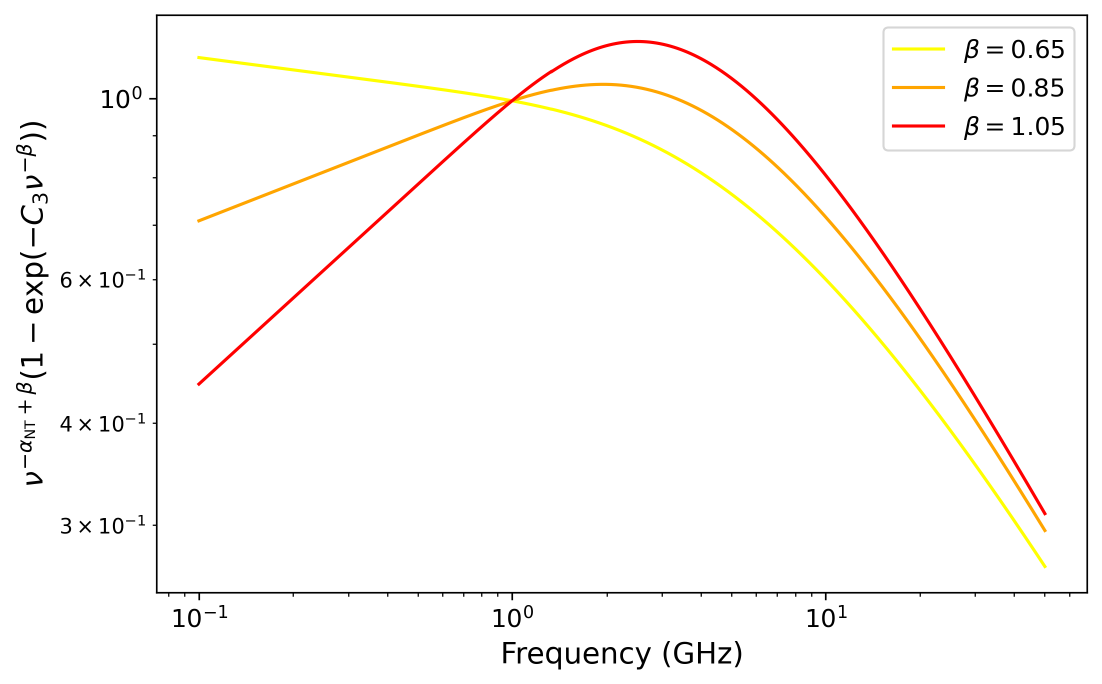}
\caption{Analysis of the behaviour of the SED as a function of $\beta$. Upper panel: absorption argument for i-FFA. Middle panel: i-FFA suppression curve. Lower panel: resulting behaviour of the synchrotron component.\label{fig:generic_beta}}
\end{figure}

Our discussion of the generic SED revealed a behaviour that deserves some additional comments. In Fig.\,\ref{fig:generic_bcde}, Sets (c) and (e) show the occurrence of a balance point close to 1\,GHz. We focus our discussion on Set (c), where $\beta$ is varied while other parameters are left at their default value (see Table\,\ref{tab:genericgrid}). To gain some insight into this topic, we first plotted what we call the absorption argument for i-FFA, which consists of the exponential argument for the attenuation by i-FFA in Eq.\,\ref{eq:generic}, i.e. it carries the physical meaning of i-FFA optical depth. The three curves displayed in the upper panel of Fig.\,\ref{fig:generic_beta} show that 1 GHz is a privileged spectral position. This is even further confirmed by the suppression curves shown in the middle panel of Fig.\,\ref{fig:generic_beta}: a flat plateau is displayed up to about 1 GHz, followed by a decay at higher frequencies. The optical depth needs to be small enough to warrant a significant departure from 1 in the suppression curves. Although the formal i-FFA turnover frequency in this set of parameters is at a few GHz, the divergence between curves as a function of $\beta$ actually starts close to 1 GHz. That specific spectral position turns out to result directly from the processing by the radiative transfer equation solution (in its shape valid for i-FFA) fed by the parameter values valid for the physical processes considered in this discussion.

The resulting behaviour is displayed in the lower panel of Fig.\,\ref{fig:generic_beta} where the incidence of the frequency factor affected by both $\alpha_\mathrm{NT}$ and $\beta$ is added. These curves are not modulated by the f-FFA factor, hence the different low frequency shape compared to Fig.\,\ref{fig:generic_bcde}. We clearly see a divergence of the three curves, with the balance point located close to 1 GHz. Below the turnover, the curves follow the $\nu^{-\alpha_\mathrm{NT} + \beta}$ trend, keeping in mind that the assumed parameter set includes $\alpha_\mathrm{NT} = 0.7$.
As a final note, we mention that the synthetic spectra proposed by \citet{Pit2} for various sets of parameters, in particular in their Fig.\,14, display a balance point that is also very close to 1\,GHz.

\end{appendix}

\end{document}